\documentclass{ws-ijmpa}
\usepackage[super,compress]{cite}
\usepackage[dvips,breaklinks]{hyperref}
\hypersetup{colorlinks,urlcolor=black,citecolor=black,linkcolor=black,filecolor=black}
\usepackage{exercise}
\usepackage{breakurl}
\def\tu{{\bar u}}
\def\tv{{\bar v}}
\def\tx{{\bar x}}

\def\trho{{\bar \rho}}

\begin{document}

\markboth{Marco O. P. Sampaio}
{Radiation from a D-dimensional collision of shock waves: numerical methods
}

%
\catchline{}{}{}{}{}
%

\title{\uppercase{Radiation from a D-dimensional collision of shock waves: numerical methods}
}

\author{MARCO O. P. SAMPAIO}

\address{Departamento de F\'isica da Universidade de Aveiro and I3N \\
Campus de Santiago, 3810-183 Aveiro, Portugal,\\
msampaio@ua.pt}

\maketitle


\begin{abstract}
We present a pedagogical introduction to the problem of evolving a head on collision of two Aichelburg-Sexl gravitational shock waves in $D$-dimensions, using perturbative techniques. We follow a constructive approach with examples, going in some detail through: the set up of the exact initial conditions and their properties; perturbative methods in flat space-time with Green function solutions; and numerical strategies to evaluate the integral solutions. We also discuss, briefly, radiation extraction methods adapted to this problem, together with some of the results for this system. 

\end{abstract}

\ccode{PACS numbers:}

\section{Introduction}	

The study of transplanckian black hole collisions in higher dimensional General Relativity (GR) has attracted a lot of attention in recent years. The motivation ranges from the possibility of probing TeV strong gravity at the LHC, to the study of the gauge gravity duality, or even to the attempt of understanding the structure of GR using $D$ as a parameter\cite{NRHEP,Emparan:2013moa,Coelho:2013zs}. 

This process, at small impact parameters, inevitably involves the formation of a horizon, whose size grows with the energy scale of the collision. It was in fact early argued by t'Hooft\cite{'tHooft:1987rb} that, in the transplanckian limit, the process should be dominated by classical general relativity and all details of the short range physics would be cloaked by a horizon\cite{Giddings:2001bu}. In particular, the details of the colliding particles should not be important in this limit, as long as they are sufficiently localised\cite{Giddings:2004xy}. This idea has been recently verified within numerical relativity in several setups, namely the collision of highly boosted black holes, boson stars and self-gravitating fluid spheres\cite{Sperhake:2008ga,Choptuik:2009ww,East:2012mb}. 

In the highly transplanckian limit, the colliding particles are greatly boosted, travelling very close to the speed of light. This has motivated the study of gravitational shock wave collisions as a model for the gravitational fields of the highly boosted particles\cite{D'Eath:1976ri,D'Eath:1992hb,D'Eath:1992hd,D'Eath:1992qu,Eardley:2002re,Rychkov:2004sf,Yoshino:2005hi,Herdeiro:2011ck,Coelho:2012sya,Coelho:2012sy}. Furthermore, also in the context of TeV gravity models\cite{ArkaniHamed:1998rs}, which predict microscopic black hole formation at the LHC for partonic centre of mass energy beyond the TeV\cite{Dimopoulos:2001hw,Giddings:2001bu,Frost:2009cf,Dai:2007ki,Sampaio:2012ei}, the colliding particles must travel at a speed $v$ very close to the speed of light ($v\gtrsim 0.999c$). Thus the study of shock wave collisions in higher dimensions provides a good model for the gravitational process also in this context. 

In sum, the study of this system is important from various perspectives: i) as a complementary method to numerical relativity for collisions at very large boost (where the numerical methods become difficult), ii) to provide better bounds on the energy lost into gravitational radiation in scenarios of TeV gravity black hole production (which are not yet very constrained\cite{Khachatryan:2010wx,Chatrchyan:2013xva,Aad:2011bw,Park:2011je}) and iii) to understand the structure of GR using $D$ as a parameter\cite{Coelho:2012sya,Emparan:2013moa}. In these lecture notes we will explore this very particular problem which, however, possesses a non-trivial structure requiring the application of many different techniques/approaches both analytic and numeric. 

From the analytic point of view it requires: i) understanding how to set up initial conditions for the evolution problem in a suitable gauge;  ii) the study of perturbation theory on a fixed background at non-linear order as to address the evolution; iii) the discussion of radiation extraction methods. These are problems which, in a way, are all common to any numerical GR problem. 

On the other hand, from the numerical point of view, it requires the discussion: i) of non-trivial integrals obtained from Green function representations; ii) of numerical strategies to compute non-trivial integration domains, and iii) the treatment of delicate numerical issues such as singularities and the use of asymptotic/series expansions to control round-off errors in difficult regions of the parameter space. Thus, the variety of problems that one is required to solve makes this problem an excellent ground to illustrate many techniques within a well defined, tractable an interesting problem. Finally, these techniques (most notably perturbation theory on a background, Green function methods and many of the numerical strategies) are certainly useful and applicable to many other relevant problems.

These lecture notes come with two companion {\em Mathematica} notebooks with examples: one with analytic calculations\cite{notebook1} and another with some numerical examples\cite{notebook2}. In the first notebook\cite{notebook1} a considerable use of some packages of the {\em xAct} suite\cite{xAct} has been made, particularly {\em xCoba}, to analyse properties of the Aichelburg-Sexl shock wave, and {\em xPert} for some derivations of perturbation theory. Throughout the notes there are exercises for the reader, with solutions found in the companion notebooks.

The structure of these notes is the following. In Sect.\,\ref{ASboostProperties} we derive the Aichelburg-Sexl solution through an ultra boost, and analyse some of its geometrical and physical properties. In Sect.\,\ref{SuperpositionASwaves} we superpose two such solutions as to collide them, discuss the properties of the superposition and formulate initial conditions for the exact evolution problem. Sect.~\ref{Sec:ApproxFlat} is dedicated to the discussion of perturbative methods in flat space and some specifics for the shock wave collision problem. In particular formal solutions are presented to all orders in perturbation theory. In Sect.~\ref{Sec:DetailedNumerics} we specialise to the shock wave collision problem, describing the gauge fixing  procedure (Sect.~\ref{subsec:gaugefix}) and presenting simplified integral expressions for the metric perturbations (Sect.~\ref{subsec:SimplifiedSols}). In Sect.~\ref{subsec:NumericalStrateg} the numerical strategies to evaluate the integral solutions are described. The results at linear order, together with a discussion of radiation extraction, are presented in Sect.~\ref{subsect:FirstOradExtract}. Finally in Sect.~\ref{subsec:highO2D} we comment on higher order results and we finish with some concluding remarks in Sect.~\ref{Sect:FinalRemarks}.
 
\section{Shock waves collision initial value problem}

\subsection{The Aichelburg-Sexl shock wave}\label{ASboostProperties}
\begin{figure}
\includegraphics[width=\linewidth]{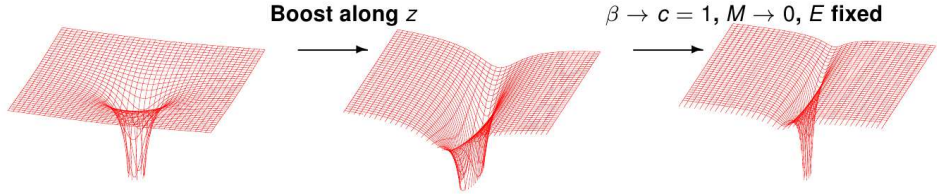}
\caption{\label{FigBoostLimit}Schematic representation of the qualitative behaviour of $-|R^{\mu\nu\alpha\beta}R_{\mu\nu\alpha\beta}|$ (Kretschmann scalar)  when the Aichelburg-Sexl ultraboost is applied to the Schwarzschild-Tangherlini black hole (horizontal directions represent the isotropic spatial coordinates $x_i$) at fixed time. In the limit of large boost, the Lorentz contraction squeezes the gravitational field on the transverse plane where the particle is centred.}
\end{figure}

As a model for the gravitational field of each colliding particle, we will use Aichelburg-Sexl (AS) shock waves~\cite{Aichelburg:1970dh}. An AS shock wave is a solution of a point-like source moving at the speed of light, thus it represents the gravitational field of a null point-like particle. It was first obtained in $D=4$ by Aichelburg and Sexl by boosting the Schwarzschild solution and taking the speed of light limit while keeping the energy parameter fixed. Let us consider $D>4$ for simplicity (see~Ref.\refcite{Aichelburg:1970dh} for the special case $D=4$). The $D$-dimensional Tangherlini solution with mass $M$ is\cite{Tangherlini:1963bw}
\begin{equation}
ds^2 = -\left[1- \tfrac{16\pi G_D M}{ (D-2) \Omega_{D-2}}\tfrac{1}{r^{D-3}}\right]
dt^2 + \left[1- \tfrac{16\pi G_D M}{(D-2) \Omega_{D-2}}\tfrac{1}{r^{D-3}}
\right]^{-1} dr^2 + r^2 d\Omega_{D-2}^2\ ,
\end{equation}
where $d\Omega_{D-2}^2$ and $\Omega_{D-2}$ are the line element and hyper-area
of the unit $D-2$ sphere, and $G_D$ is the $D$-dimensional Newton constant. To obtain the $D$-dimensional AS metric first we change to isotropic coordinates. We transform to a new radial coordinate $R$ and keep the angular coordinates:
\begin{equation}
r=R\left(1+A\right)^{\frac{2}{D-3}} \;,\;\; A\equiv \frac{1}{4}\left(\frac{r_s}{R}\right)^{D-3}\;,\;\; r_s^{D-3}\equiv \frac{16\pi G_D M}{ (D-2) \Omega_{D-2}}
\end{equation}
where $r_s$ is the Schwarzschild radius. Then ($R=\sum_{i=1}^{D-1} x_i^2$) 
\begin{equation}\label{eq:boosted_isotropic}
ds^2=-\left(\dfrac{1-A}{1+A}\right)^2dt^2+(1+A)^{\frac{4}{D-3}}\sum_{i=1}^{D-1}dx_i^2
\end{equation}
Next, we pick a spatial direction (say along $x^{D-1}\equiv z$) and perform a boost with velocity $\beta$, which amounts to the following coordinate transformation\footnote{For notational simplicity we use the same names for the transformed coordinates and apply a replacement rule notation $\rightarrow$, where the old coordinates are replaced by their expressions in terms of the new coordinates.}:
\begin{eqnarray}
t & \rightarrow & \gamma(t-\beta z) \nonumber\\
z & \rightarrow & \gamma(z-\beta t) \;\;\qquad\qquad,\qquad \gamma\equiv (1-\beta^2)^{-\frac{1}{2}}\\
 x_T & \rightarrow & x_T \equiv (x_1,\ldots,x_{D-2})\;, \qquad\rho^2\equiv \sum_{i=1}^{D-2}x_i^2\; .\nonumber
\end{eqnarray}
With this transformation, we note first that fixing the energy $E \equiv \gamma M$ and taking the limit $\beta\rightarrow 1,M\rightarrow 0 $
\begin{equation}
A=\dfrac{4\pi G_D E}{(D-2)\Omega_{D-2}}\dfrac{\sqrt{1-\beta^2}}{\left[\left(\tfrac{z-\beta t}{\sqrt{1-\beta^2}}\right)^2+\rho^2\right]^{\frac{D-3}{2}}}\Rightarrow \lim_{\beta\rightarrow 1}{A}=0\; .
\end{equation}
Then the line element in such limit is expanded as
\begin{equation}\label{eq:boosted_isotropic_boost}
ds^2=-dt^2+dz^2+dx_T^2+\tfrac{4(D-2)}{(D-3)}\frac{A}{1-\beta^2}(dt-\beta dz)^2+\ldots
\end{equation}
where $\ldots$ denotes terms that go to zero in the limit.  An intuitive schematic picture of this limiting procedure is presented in Fig.~\ref{FigBoostLimit}.
In a frame where the black hole is moving with constant velocity $\beta$, the metric is Lorentz contracted along $z$. Thus, the curved region becomes increasingly concentrated on a plane, perpendicular to the longitudinal $z$ direction, whereas the transverse directions are not transformed. 

One can check (by taking the limit $\beta\rightarrow 1$) that in fact the term proportional to $A$ in Eq.~\eqref{eq:boosted_isotropic_boost}  goes to zero off the plane $z=t$ (which moves at the speed of light) and diverges on the plane. So the limiting line element is simply flat Minkowski space off the moving plane, plus a Dirac delta distributional source on the moving plane. The proof of the limit is presented in~\ref{AppBoostLimit}, and the final line element for a particle moving in the $+z$ direction is
\begin{equation}
ds^2 = -du dv + d\rho^{2} + \rho^2 d {\Omega}^2_{D-3}+\kappa \Phi(\rho) \delta(u) du^2\ ,\label{AiSe}
\end{equation}
where $\kappa\equiv 8\pi G_D E/\Omega_{D-3}$, and $u=t-z$, $v=t+z$ are null coordinates. The function $\Phi$ depends only on $\rho$ and takes the form\cite{Eardley:2002re}
\begin{equation}
\Phi(\rho)=\left\{
\begin{array}{ll}
 -2\ln(\rho)\ , &  D=4\  \vspace{2mm}\\
\displaystyle{ \frac{2}{(D-4)\rho^{D-4}}}\ , & D>4\ \label{phidef}
\end{array} \right. \ .
\end{equation}
Clearly, a shock wave solution moving in the opposite $-z$ direction with the same energy is obtained by replacing $z\leftrightarrow-z$ or equivalently exchanging $u$ and $v$.

\subsubsection{Properties of the solution}
Some obvious properties of the solution are:
\begin{itemize}
\item {\em Axial symmetry}: The solution is invariant under rotations $d\Omega_{D-3}$ on the transverse plane;
\item {\em $v$-translations symmetry}: The metric components are independent of $v$;
\item {\em Transformation under boosts}: Boosting the solution along $+z$ with velocity $\beta=\tanh \alpha$, amounts to a rescaling $E\rightarrow e^\alpha E$ (as expected from the 4-momentum transformation of a null particle).
\end{itemize}
One can understand better the gravitational properties of the solution by analysing the curvature and the Einstein tensor. This can be done analytically for all $D$. 
\begin{Exercise}
The results of this exercise can be checked with {\em Mathematica}, using {\em xCoba} (see first companion notebook\cite{notebook1}).
\Question
Show that the only independent non-zero components of the Riemann tensor for the AS shock wave~\eqref{AiSe} are (derive the right hand side):
\begin{equation}
R_{uiuj}=-\kappa\dfrac{\delta(u)}{2}\left[\delta_{ij}\dfrac{\Phi'}{\rho}+\Gamma_i\Gamma_j\left(\nabla^2\Phi-(D-2)\dfrac{\Phi'}{\rho}\right)\right]\;, \label{RiemannAS}
\end{equation}
where $\Gamma_i\equiv x_i/\rho$  and $\nabla^2$ is the Laplacian, both on the transverse plane.
\Question
Compute the Einstein tensor and show that the only non-zero component gives the following energy-momentum tensor
\begin{equation}
T_{uu}=-\dfrac{E}{2\Omega_{D-3}}\delta(u)\nabla^2\Phi=E\delta(u)\delta^{(D-2)}(x_i) \; , \label{TuuAS}
\end{equation}
where the last step can be shown using Gauss's law on the integral $\int_Vd^{D-2}x \nabla^2\Phi$ over the volume of a $(D-2)$-sphere.
\end{Exercise}
 An alternative way of deriving these results is by computing the tensors directly in the original black hole space-time, boosting and then taking the limit.

\begin{figure}
\includegraphics[width=\linewidth]{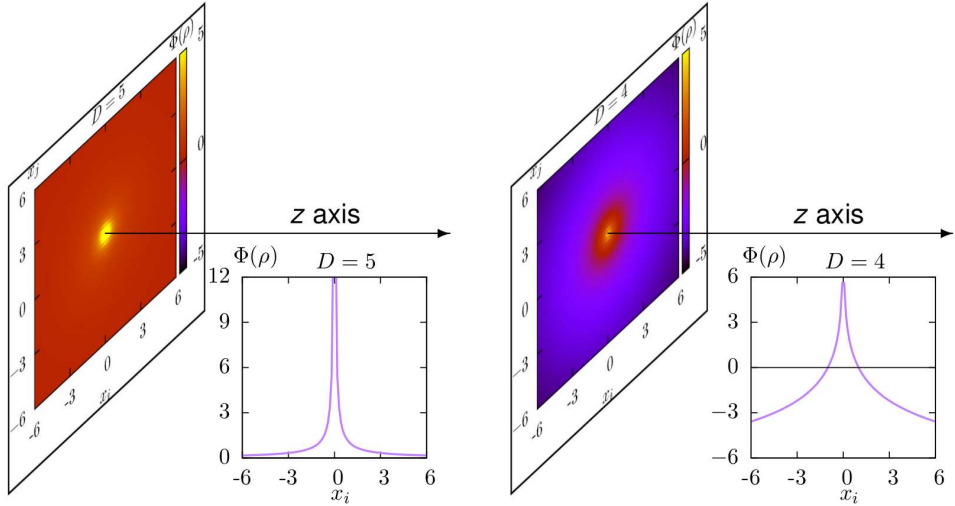}
\caption{\label{FigProfilesColor}Transverse plane profile functions for $D=5$ (left) and $D=4$ (right). Here we represent the transverse plane in perspective and the $z$-axis along which the plane moves at the speed of light. The inset plots show $\Phi$ as a function of a transverse direction $x_i$.}
\end{figure}

It is now clear from Eq.~\eqref{TuuAS} that this solution describes the gravitational field of a point-like particle moving at the speed of light with energy density distribution $T_{uu}$, and energy $E$. Looking at the Riemann tensor Eq.~\eqref{RiemannAS} we also conclude that the gravitational field is zero off the transverse plane moving at the speed of light, and has support on the whole plane. On the transverse plane the gravitational field decays away from the centre as an inverse power of $\rho$ and it is impulsive in nature due to the $\delta(u)$ distribution. This can be better visualised in Fig.~\ref{FigProfilesColor} where we show plots of the profile functions on the transverse plane.
 The profile function for $D=5$ ($D>4$ in general) decays for large $\rho$, whereas the $D=4$ special case goes to $-\infty$. This special feature is not incompatible with the considerations above, since $\Phi$ is not a gauge invariant quantity. Components of gauge invariant objects always contain at least one derivative $\Phi'$ as observed in the Riemann tensor Eq.\eqref{RiemannAS}, so the gravitational field does indeed decay away from the centre. 

Regarding the impulsive nature of the shock wave along $u$, one can see it as an idealisation of a realistic distribution where the $\delta(u)$ (as well as the $\delta^{(D-2)}(x_i)$) is replaced by a smeared out form, amounting to a matter distribution with some extension (instead of being point-like). We expect, however, that the gravitational field away from the centre of the source does not depend on the details as long as the source is sufficiently localised.

\subsubsection{Rays analysis}
Another intuitive way of understanding the AS solution is by looking at null rays (geodesics) incident on the shock plane, which suffer the gravitational deflection and redshift that test null particles do. In the next section, such null rays will also be useful to define the points which are causally connected to the collision plane, when two shock waves are collided. 

\begin{figure}
\includegraphics[width=0.335\linewidth]{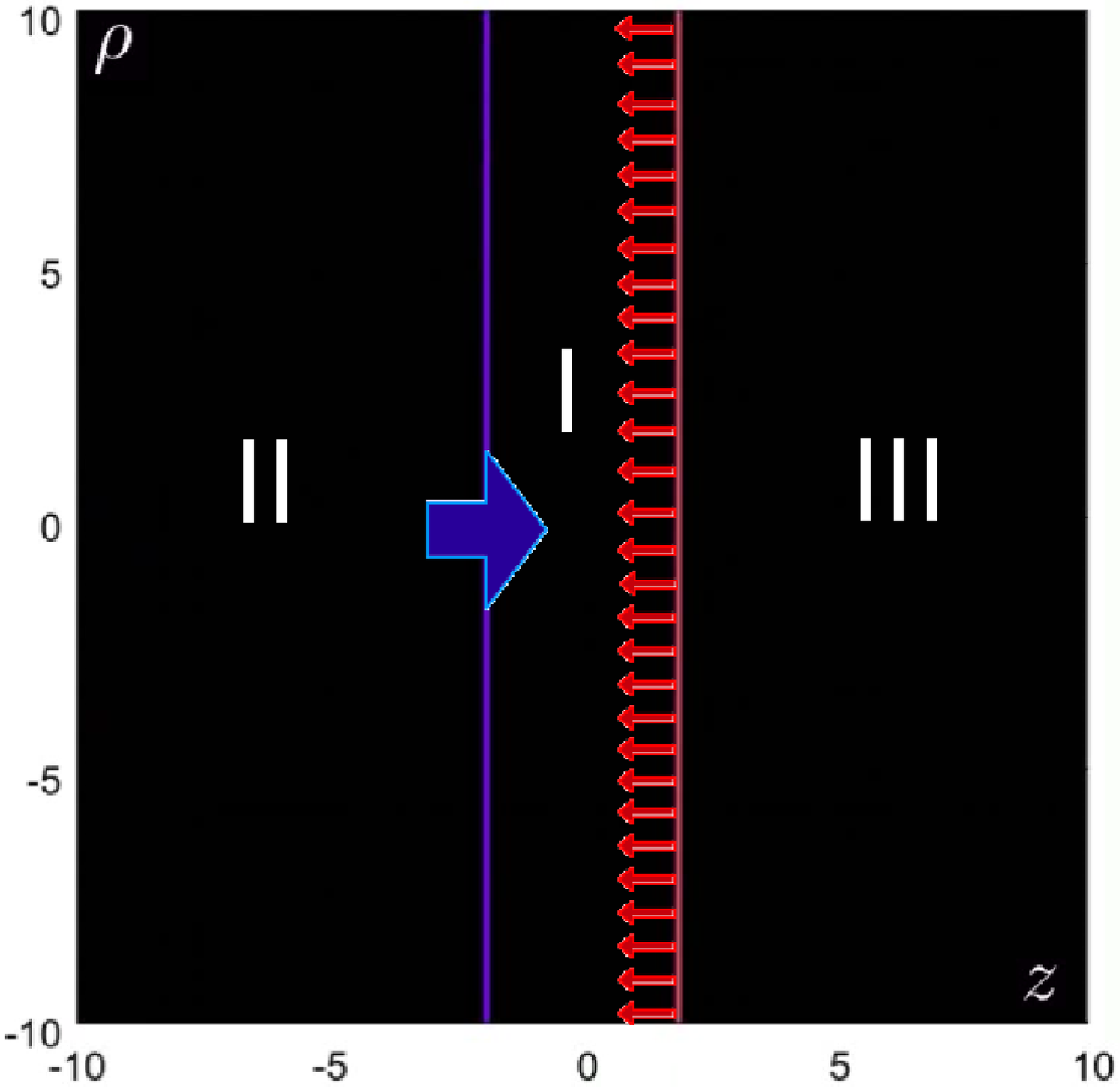}\includegraphics[width=0.336\linewidth]{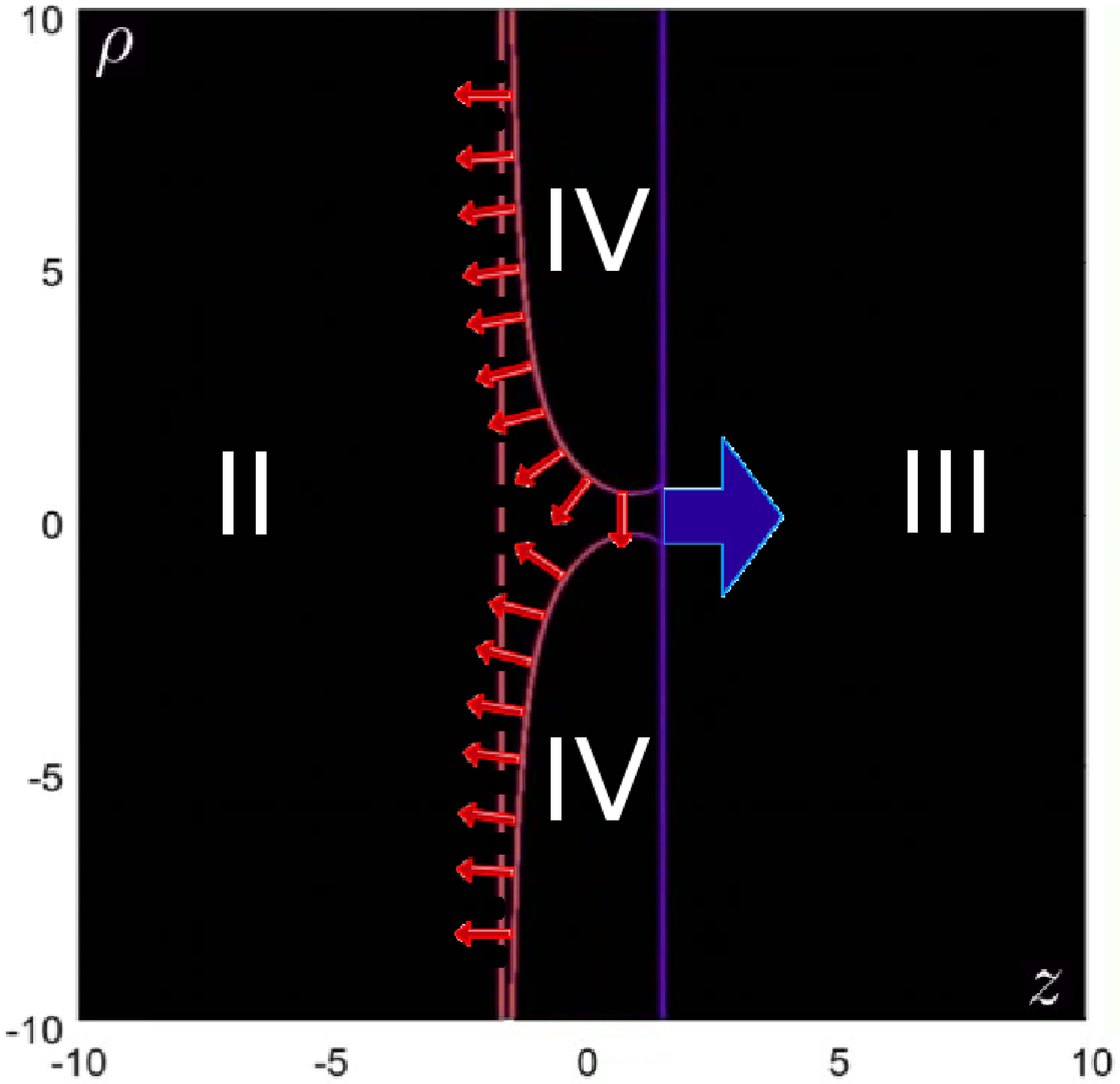}\includegraphics[width=0.335\linewidth]{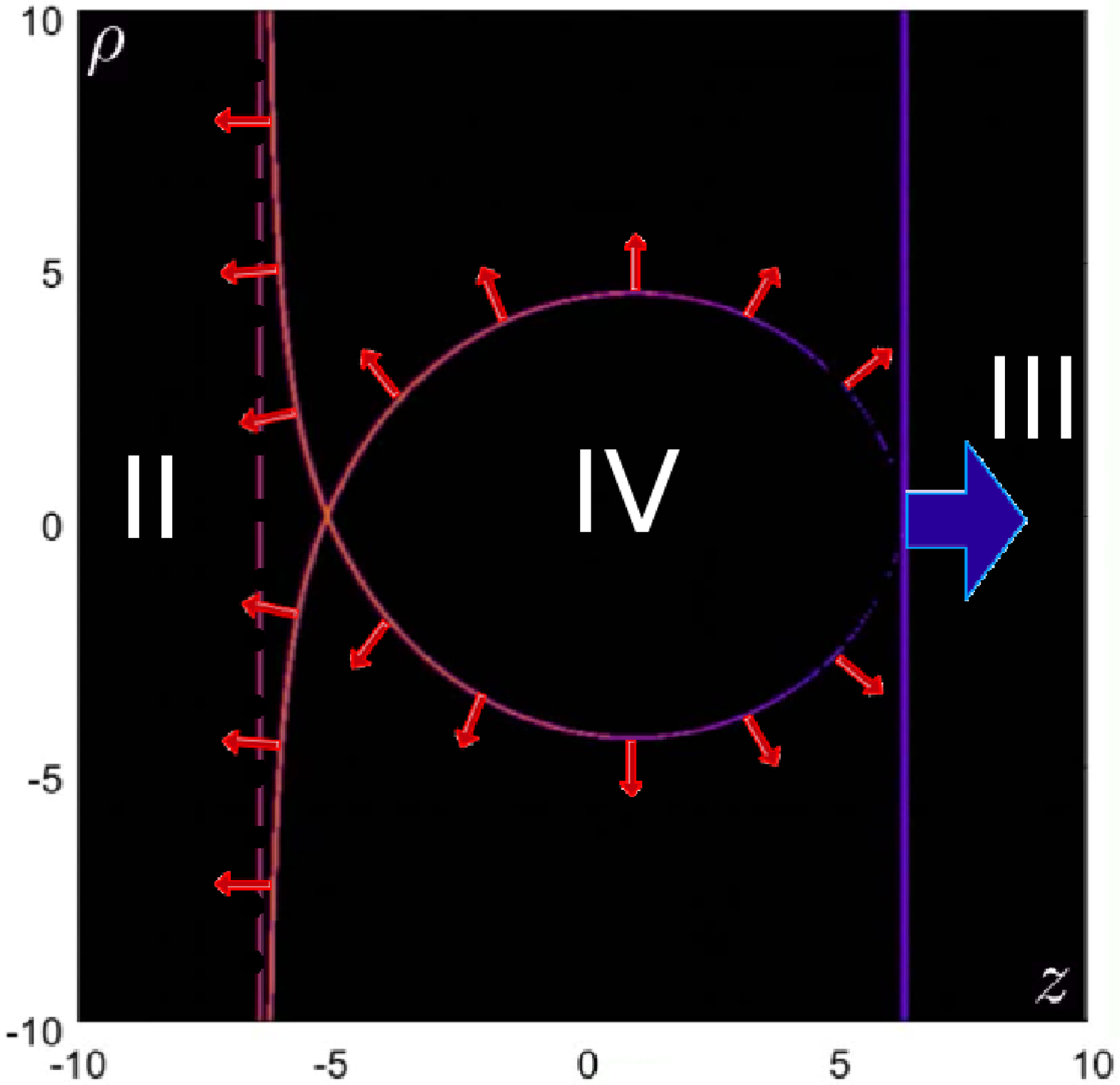}
\caption{\label{FigRaysMovie1_5D} Representation of a test wave front of null rays (small red arrows) scattering through the transverse plane of the AS shock wave travelling along the positive $z$ direction (big blue arrow), for $D=5$: i) before scattering (left), ii) a small time after scattering (middle) and iii) a later time after scattering (right). The dashed red line (right) represents the plane wave front the rays would have followed if no scattering had occurred.}
\end{figure}

Let us analyse Fig.~\ref{FigRaysMovie1_5D} where the scattered trajectories of a plane of null rays are represented (along the small red arrows). Such plane of rays is incident perpendicularly to the right moving AS shock wave (along big blue arrow) -- see left panel. 
In the middle figure we observe a snapshot of the rays, some small time after they are scattered through the shock wave, and the right panel at another later time. The rays propagate freely in flat space both before and after the scattering. The scattering event occurs instantaneously at $t=0$ when the two planes meet. The scattered rays suffer two effects. First, it is clear from the middle and right panel that the tangent vectors of all rays are bent inwards towards the collision axis, as expected from the gravitational attraction of the source at the centre. In particular the rays are less bent further away from the centre. At late times (as seen in the right panel), there is an increasingly circular envelope of rays around the scattering centre, and another outermost one which asymptotes to the dashed line representing the initial plane of rays extrapolated to the left. 

A second effect is more subtle, but equally interpretable physically. In the middle panel we observe that there are scattered rays which have not yet emerged from the central region of the shock wave, which are later identifiable in the right panel. This is due to a redshift effect suffered by all rays, which intuitively get stuck on the shock wave for some time until they emerge later in their bent trajectory. This time step is larger for rays incident close to the centre and small away from the centre, as expected, since the gravitational field decays with an inverse power of $\rho$. 
\begin{Exercise}\label{ExASgeodesics}
Consider the propagation of null geodesics on the background~\eqref{AiSe} described by the action 
\begin{equation}S=\dfrac{1}{2}\int d\lambda\, g_{\mu\nu}\dfrac{dx^\mu}{d\lambda}\dfrac{dx^\nu}{d\lambda}\equiv\dfrac{1}{2}\int d\lambda\, g_{\mu\nu}p^\mu p^\nu
\end{equation}
($x^{\mu}(\lambda)$ where $\lambda$ is an affine parameter).
\Question
Use the conservation law for $p^v$ to show that we can choose the parameter to be $\lambda=u$.
\Question
Use the Euler-Lagrange equations for $x^i(\lambda)$ (transverse plane coordinates) to show that
\begin{equation}\label{EqASgeoX}
x^i=x_0^i+{x'}_0^i\, u+\dfrac{\kappa}{2}\nabla_i\Phi(\rho_0)u\,\theta(u)
\end{equation}
where $x_0^i$ and ${x'}_0^i$ are integration constants and $\theta(u)$ is the Heaviside step function. (Note: ${x'}_0^i=0$ for rays incident perpendicular to the shock plane).
\Question \label{ExASgeodesics3}
Use the conserved Hamiltonian for null rays, $p_\mu  p^\mu=0$, to show that
\begin{equation}\label{EqASgeoV}
v=v_0+\kappa\Phi(\rho_0)\theta(u)+u\sum_i\left[{x'}_0^i+\dfrac{\kappa}{2}\nabla_i\Phi(\rho_0)\,\theta(u)\right]^2 \; ,
\end{equation}
where $v_0$ is an integration constant.
\end{Exercise}

\subsection{Superposition principle \& Apparent horizon}
\label{SuperpositionASwaves}
In the previous section we have seen that an AS solution moving along $z=t$ represents a massless particle moving at the speed of light, and that its gravitational field is restricted to the transverse plane. We also saw that an incident ray, is not affected by the shock wave before it crosses the transverse plane. This is simply due to causality, i.e. no signal of the shock wave can reach the incident ray faster than light, in particular the incident ray cannot feel any gravitational field before the crossing. This causality argument allows us to superpose another AS shock wave moving in the opposite direction along $z=-t$, and we immediately know the exact solution corresponding to the two waves before collision, i.e. it is given by the superposition of the two line elements.  

\subsubsection{Rosen coordinates}
In Exercise~\ref{ExASgeodesics}.\ref{ExASgeodesics3} we have seen that in the coordinates we have used so far (designated of Brinkmann type\cite{Brinkmann}), the geodesics and their tangent vectors are discontinuous across the shock wave (see Eqs.~\eqref{EqASgeoX} and~\eqref{EqASgeoV}). This makes it difficult to identify the collision event when superposing two shock waves, because there should be discontinuities of one wave crossing the other and vice-versa. Thus, before performing the superposition, it is convenient to change to a chart where the coordinates are continuous. The new coordinate system (called Rosen  coordinates\cite{Rosen}) is defined by\cite{Eardley:2002re}
\begin{eqnarray}
u &=& \tu\ ,\nonumber \\
v &=& \tv+\kappa\theta(\tu)\left(\Phi + \frac{\kappa \tu (\bar{\nabla}\Phi)^2}{4}\right)\ = \tv+\kappa\theta(\tu)\left(\Phi + \frac{\kappa \tu \Phi'^2}{4}\right),\nonumber \\
x^i&=& \tx^i + \kappa\frac{\tu}{2} \bar{\nabla}_i \Phi(\tx)\theta(\tu) \Rightarrow \left\{\begin{array}{rcl} \rho&=& \trho\Big(1+ \frac{\kappa \tu \, \theta(\tu)}{2 \bar \rho}\Phi'\Big)  \\
\phi_{a}&=&\bar \phi_a  \end{array}\right.
\ , \label{ct}
\end{eqnarray}
where $\Phi$ and its derivative $\Phi'$ are evaluated at $\bar\rho$, $\phi_a$ are the angles on the $(D-3)$-sphere and $a=1...D-3$.
\begin{Exercise}
Show that the new coordinates~\eqref{ct} correspond to the parameters of the null rays of Exercise~\ref{ExASgeodesics}, which are incident perpendicularly to the shock plane, i.e. with the replacements $\lambda\rightarrow\tu$, $x^i_0\rightarrow\tx^i$, $v_0\rightarrow\tv$ and ${x'}_0^i=0$.
\end{Exercise}
In Rosen coordinates~\eqref{ct} the metric of one AS shock wave becomes\cite{D'Eath:1976ri,Rychkov:2004sf,Yoshino:2005hi}
\begin{equation}
ds^2 = -d\tu d\tv + \Big(1+\dfrac{\kappa \tu \theta(\tu)}{2}\Phi''\Big)^2d\trho^{2} + \trho^2\Big(1+ \frac{\kappa \tu \, \theta(\tu)}{2 \bar \rho}\Phi'\Big)^2 d\bar{\Omega}^2_{D-3} \ .\label{newmetric}
\end{equation}
An identical shock wave travelling in the  $-z$ direction is obtained by changing $z\rightarrow -z$ in~\eqref{AiSe} or equivalently by exchanging $\bar{v}\leftrightarrow \bar{u}$ in~\eqref{newmetric}. The superposition of two such waves describes the space-time of the collision everywhere except in the future light cone of the collision. The superposed line element is\footnote{Here we have used the identities $d\trho^2=\bar{\Gamma}_i\bar{\Gamma}_jd\tx^id\tx^j $ and $\trho^2d\bar{\Omega}^2_{D-3}=(\delta_{ij}-\bar{\Gamma}_i\bar{\Gamma}_j)d\tx^id\tx^j$.}
\begin{multline}ds^2 = -d\tu d\tv +\delta_{ij}d\tx^id\tx^j + \left[2\left(\dfrac{\kappa\Phi'}{2\trho}\right)\left(\tu\theta(\tu)+\tv\theta(\tv)\right)\bar{\Delta}_{ij}+\phantom{\left(\dfrac{\kappa}{\trho^{D-2}}\right)^2}\right. \\\left.+\left(\dfrac{\kappa\Phi'}{2\trho}\right)^2\left(\tu^2\theta(\tu)+\tv^2\theta(\tv)\right)\left((D-3)\delta_{ij}-(D-4)\bar{\Delta}_{ij}\right)\right]d\tx^id\tx^j 
 \ ,\label{collision}
\end{multline}
 where  $\bar{\Delta}_{ij}\equiv \delta_{ij}-(D-2)\bar{\Gamma}_i\bar{\Gamma}_j$ is a traceless tensor on the transverse plane with $\bar{\Gamma}_i=\tx_i/\trho$, angular factors. 
\begin{figure}
\begin{center}
\mbox{\includegraphics[width=0.49\linewidth]{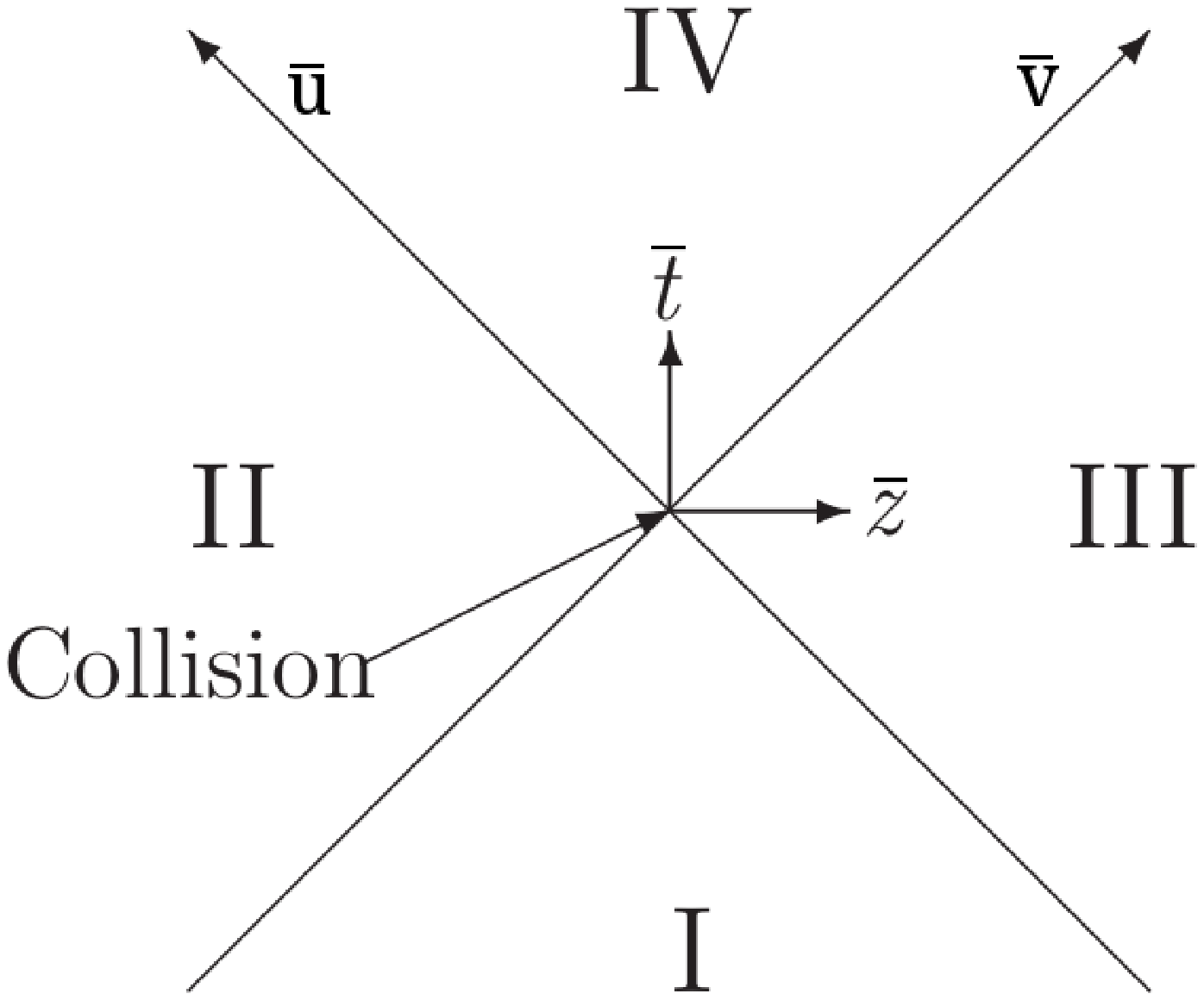}\hspace{0.02\linewidth}\includegraphics[width=0.49\linewidth]{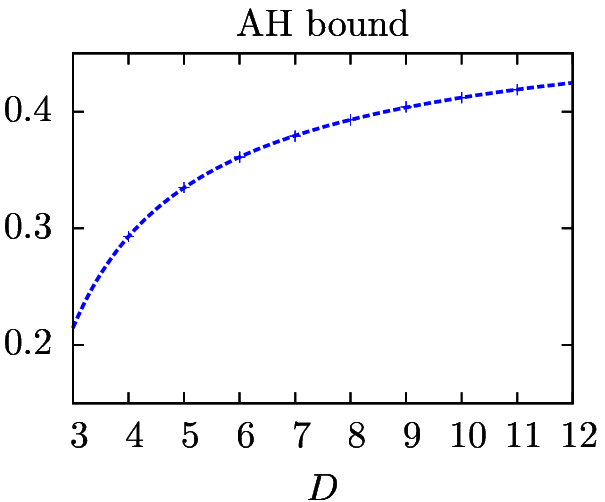}}
\end{center}
\caption{\label{barredDiagAH} Space-time diagram of the collision (left) and apparent horizon upper bound on the inelasticity (right) as a function of $D$.}
\end{figure}
The left diagram of Fig.~\eqref{barredDiagAH} shows a projected space-time diagram in Rosen coordinates, where the transverse directions have been suppressed. The two shock waves travel along $\tu=0$ and $\tv=0$ respectively, and they  collide at $\tu=\tv=0$. In this chart one can easily identify the future light cone of the collision as the space-time region $\tu>0\wedge\tv>0$ (region IV), by extending the coordinates continuously across $\tu=0$ and $\tv=0$, since there is no discontinuity in the null geodesics or their tangent vectors as we cross the shock waves. The regions I, II and III are flat and correspond to the regions: i) between the shock waves before collision, ii) behind the $\tu$-shock and iii) behind the $\tv$-shock, respectively. We have also written the metric in a suggestive form separating a first part that looks like the usual Minkowski line element, and the other terms with the $\trho$ dependence factored out in a power of $\kappa\Phi'/(2\trho)=\kappa/\trho^{D-2}$ which suppresses those terms for large $\trho\gg \kappa^{1/(D-3)}$, with $\tu,\tv$ fixed. 

Finally, another central property of this exact superposition is the existence of an apparent horizon on the null surface composed by the union of the surfaces $\tu=0\wedge \tv<0$ and $\tv=0\wedge \tu<0$ which was found first by Penrose in four space-time dimensions and later generalised to $D$ dimensions by Eardley and Giddings\cite{Eardley:2002re}. The apparent horizon area results on an upper bound on the fraction of the centre of mass energy, $\epsilon_{\rm radiated}$ (inelasticity), which is radiated away in a head on collision. The bound increases with dimension and is given by
\begin{equation}
\epsilon_{\rm radiated}\le 1-\frac{1}{2}\left(\frac{D-2}{2}\frac{\Omega_{D-2}}{\Omega_{D-3}}\right)^{\frac{1}{D-2}} \ .
\end{equation} This is plotted in the left panel of Fig.~\ref{barredDiagAH} where we can see it increases monotonically with dimension approaching $50\%$ in the limit of infinite $D$. The existence of an apparent horizon which traps the colliding sources, further supports the independence of the results on the details of the localised source. In particular the fact that our colliding particles are point-like, instead of spread out over a small volume, is not expected to affect black hole formation  in the transplanckian limit\cite{Giddings:2004xy,Sperhake:2008ga,Choptuik:2009ww,East:2012mb}.

\subsubsection{Coordinates adapted to one shock wave \& exact initial conditions}

As already mentioned, the form of the metric in Eq.~\eqref{collision} is exact except in the future light cone of $\tu=\tv=0$. The evolution of the initial data at $\tu=0$ and $\tv=0$, given by Eq.~\eqref{collision}, consists of a complicated non-linear problem in solving Eistein's equation in region IV. However, as already noted, the metric resembles a perturbation of flat space in the usual Minkowski coordinates, so an approximation scheme seems possible. The regions where the metric looks like a perturbation are for 
$\trho\gg (\kappa\tu)^{1/(D-2)}$ and  $\trho\gg (\kappa\tv)^{1/(D-2)}$ (this can be obtained by requiring that the $\left\{\tu,\tv,\trho\right\}$ dependent terms inside square brackets in Eq.~\eqref{collision} are small). 
The main shortcoming of Rosen coordinates in formulating an approximation is related to these two conditions. They imply that, when going into regions II or III where the metric is {\em exactly} Minkowski space, its form is not in the standard Minkowski coordinates and depends on the shock waves parameter $\kappa$. However, after the superposition, we can partially fix this shortcoming by returning to new Brinkmann type coordinates which are adapted to one of the shock waves (say the $+z$ moving one).  This is obtained in the next exercise by undoing the transformation in Eq.~\eqref{ct}, but now, because of the other shock wave, we obtain extra terms.
\begin{Exercise}\label{Ex4}
\Question
Using transformation~\eqref{ct}, show that 
\begin{equation}
d\tx^j\left(\delta^i_j+\dfrac{\kappa \Phi'(\trho)}{2\trho^{D-3}}\dfrac{u\theta(u)}{\trho}\bar{\Delta}^i_j\right)\equiv M^i_jd\tx^j=dx^i-\dfrac{\kappa \Phi'(\trho)\theta(u)}{2}\bar{\Gamma}^idu \label{invertct}
\end{equation}
\Question\label{EX4.2}
Show that the inverse is in the form $(M^{-1})^i_j=A\delta^i_j+B\bar{\Delta}^i_j$ and that on $u=0^+$, $A=1,B=0$ so $d\tx^j=dx^j-\kappa \Phi'(\rho)\Gamma^j(\rho) du/2$.
\Question
Convince yourself, by splitting the $u$-shock part from the $v$-shock part of~\eqref{collision} that in the adapted Brinkmann coordinates
\begin{multline}ds^2 = -du dv +\delta_{ij}dx^idx^j+\kappa \Phi(\rho) \delta(u) du^2+\\ \left\{-2\bar{h}(u,v,\rho)\bar{\Delta}_{ij}+\bar{h}(u,v,\rho)^2\left((D-3)\delta_{ij}-(D-4)\bar{\Delta}_{ij}\right)\right\}d\tx^id\tx^j 
 \ ,\label{collisionBrink}
\end{multline}
with $d\tx^i$ replaced using the inversion of Eq.~\eqref{invertct} and $\tv,\trho$ replaced using the inversion of~\eqref{ct}, and where
\begin{equation}
\bar{h}(u,v,\rho)=-\dfrac{\kappa\Phi'\tv}{2\trho}\theta(\tv) \; .
\end{equation}
\end{Exercise}

In Exercise~\ref{Ex4}, Eq.~\eqref{collisionBrink} we have shown that by returning to Brikmann coordinates adapted to one of the shock waves, the line element looks like a $u$-shock background (which is flat space in the usual Minkowski coordinates plus the impulsive term), and a contribution from the $v$-shock. The latter can be regarded as a perturbation as long as $\bar{h}(u,v,\rho)\ll 1$. This separation will turn out to be useful, but for now we can formulate exact initial condition for the collision event in these adapted Brinkmann coordinates. Before proceeding it is convenient to use a new system of natural units where $\kappa=1$. We also make a conventional change in the normalisation of $u,v$ by a factor of $\sqrt{2}$, so the overall rescaling to natural units is $(u,v,x^i)\rightarrow \kappa^{1/(D-3)} (\sqrt{2}u,\sqrt{2}v,x^i)$.

To visualise better the setup, it is very useful to represent it in a space-time diagram. First let us note that from the point of view of the $u$-shock, the null rays travelling on the plane of the $v$-shock (i.e. its generators) will scatter through the $u$-shock exactly as the test incident rays of Fig.~\ref{FigRaysMovie1_5D}\footnote{This is because we are using Brinkmann coordinates adapted to the $u$-shock.}. So the outermost envelope or rays in Fig.~\ref{FigRaysMovie1_5D}, defines the causal separation between the curved region (IV) and the flat region (II) after the $v$-shock null generators. This is also valid for the superposition, and the only unknown part is region IV. 
\begin{figure}
\begin{center}
\includegraphics[width=\linewidth]{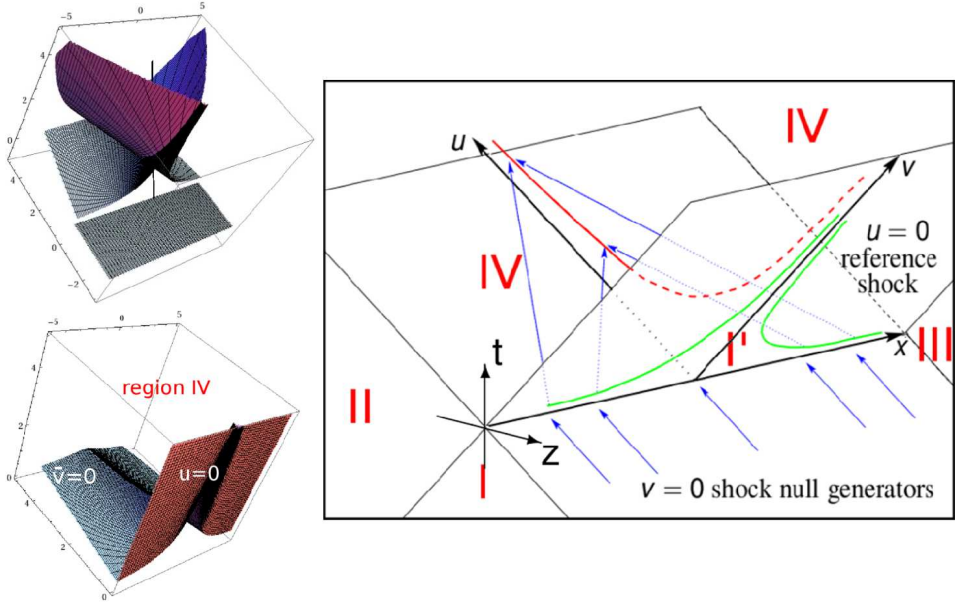}
\end{center}
\caption{\label{5DspacetimeDiag} Space-time diagrams in adapted Brinkmann coordinates. {\em Right (adapted from~\cite{Herdeiro:2011ck}):} Detailed diagram with identification of axis ($t,z,\rho\equiv x$), regions (red symbols), $v$-shock generators (blue) and the collision (in green). {\em Top left:} Surface defined by the null generators of the $v$-shock as they scatter through the $u$-shock. {\em Bottom left:} Causal surfaces ($u=0\wedge v>0$ and $\tv=0\wedge u>0$) separating the future light cone of the collision, region IV.}
\end{figure}
In Fig.~\ref{5DspacetimeDiag}, on the right, we represent a space-time diagram of the superposition in $u$-adapted Brinkmann coordinates. The $u$-shock rays travel along $u=0$ without any discontinuity, however the $v$-shock generators, defined by $\tv=0$, travel along  $v=0$  for $u<0$ (see blue incident arrows in the diagram), but on $u=0$ they suffer a jump to the green line (the collision) and are focused towards the axis where they cross (see blue arrows going into region IV). In fact, the green line is defined by the condition
\begin{equation}
 \tu=\tv=0\Rightarrow u=0\; ,\; v=\dfrac{\Phi(\rho)}{\sqrt{2}} \; .
\end{equation}
 The top left 3D diagram shows the surface defined by the $v$-shock generators more clearly where, for comparison, the world-line of a time-like observer is represented (in black). In the bottom left plot we have removed the extrapolation of the $v$-shock null rays into the curved region (see top plot) and have coloured in red the null rays of the $u$-shock ($u=0$ plane in the bottom plot), for an easy visualisation of region IV\footnote{Observe that in this coordinate system, there is an extra flat region just below region IV (denoted region $I'$ in the right diagram of Fig.~\ref{5DspacetimeDiag}) which is due to the jump of the $v$-shock rays to the green line.}. The blue ($u=0$) and red ($\tv=0$) surfaces are thus the boundary of the unknown curved region IV. It is straightforward to observe from~\eqref{collisionBrink} that on the blue surface ($\tv=0 \wedge u>0$), the line element is that of Minkowski space in standard coordinates
\begin{equation}
g_{\mu\nu}(u>0,\tv=0,x_i)=\eta_{\mu\nu} \; .\label{InitCond1}
\end{equation}
 On the other hand, on $u=0^+$ (red surface), the metric takes the form (exactly)
\begin{equation}
g_{\mu\nu}(u=0^+,v,x_i)\equiv\eta_{\mu\nu}+h_{\mu\nu}=\eta_{\mu\nu}+h_{\mu\nu}^{(1)}+h_{\mu\nu}^{(2)}\label{InitCond2}
\end{equation}
where $h_{\mu\nu}^{(i)}$ can be read out directly from~\eqref{collisionBrink} using the results of Exercise~\ref{Ex4}.\ref{EX4.2}, specialised on $u=0^+$:
\begin{multline}\label{InitCondRead}
  h_{\mu\nu}dx^\mu dx^\nu = (D-3)\Phi'(2h(v,\rho)+(D-3)h(v,\rho)^2)\left[\dfrac{\Phi'}{2}du^2-\sqrt{2}\Gamma_idx^idu\right] \\ + \left(-2h(v,\rho)\Delta_{ij}+h(v,\rho)^2\left((D-3)\delta_{ij}-(D-4)\Delta_{ij}\right)\right)dx^idx^j 
\end{multline}
with 
\begin{equation}\label{Inithvrho}
h(v,\rho)\equiv\bar{h}(0,v,\rho)=-\dfrac{\Phi'}{2\rho}(\sqrt{2}v-\Phi(\rho))\theta(\sqrt{2}v-\Phi(\rho)) \; .
\end{equation}
The order $n$ of $h_{\mu\nu}^{(n)}$ is defined by the power $h(v,\rho)^n$ appearing in corresponding terms of $h_{\mu\nu}$. The non-trivial conditions, Eq.~\eqref{InitCondRead}, on $u=0^+$ encode the information on the scattering of the null rays of the $v$-shock during the collision, in coordinates where the $u$-shock rays are unaffected. In a way it amounts to adopting the reference frame of the $u$-shock. It is important to note that the initial conditions are however {\em exact}, and all that was done so far was to use a convenient gauge for the superposition. 

Finally, the exact initial value problem amounts to solving Einstein's equations with the initial conditions~\eqref{InitCond1} and~\eqref{InitCond2}. Regarding~\eqref{InitCond1}, it is trivially obeyed once condition~\eqref{InitCond2} is imposed on $u=0^+$, since as we move below the blue surface (see Fig.~\ref{5DspacetimeDiag}) the propagation of the initial conditions~\eqref{InitCond2} from $u=0^+$ in flat space\footnote{This is the correct procedure since the blue surface is the causal null boundary with flat space, so the curvature must drop to zero.} also ceases to have support, so we consistently, obtain flat space in the usual Minkowski coordinates.

\section{Approximate solution \& Perturbative methods in flat space}
 \label{Sec:ApproxFlat}
In the previous section we have setup the exact initial value problem for the collisions of two shock waves with the same energy parameter in adapted Brinkmann coordinates. The exact initial conditions on the $u=0^+$ surface, Eq.~\eqref{InitCond2} and~\eqref{InitCondRead} take a form which resembles a perturbation of flat Minkowski space. This, however, is only true in regions 
\begin{equation}
|h(v,\rho)|\ll 1 \Rightarrow g(\rho)\equiv \left|\dfrac{\Phi(\rho)}{\sqrt{2}}+\dfrac{\rho}{2\sqrt{2}\Phi'(\rho)}\right|\gg v
\end{equation}
on $u=0^+$. 
\begin{figure}
\includegraphics[width=0.34\linewidth,height=0.35\linewidth]{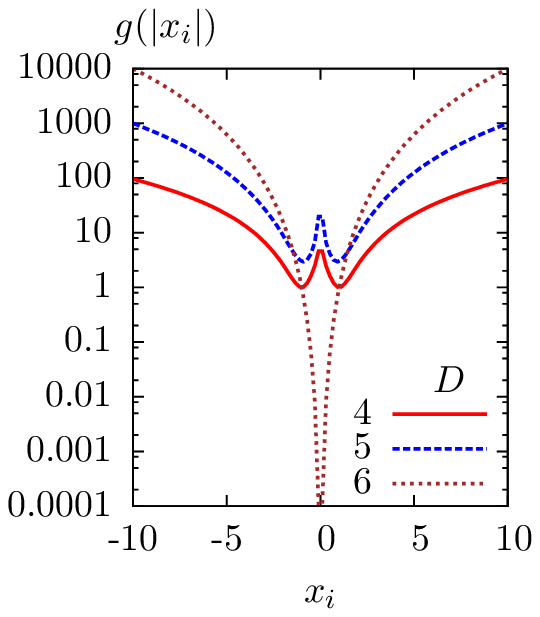}\includegraphics[width=0.35\linewidth]{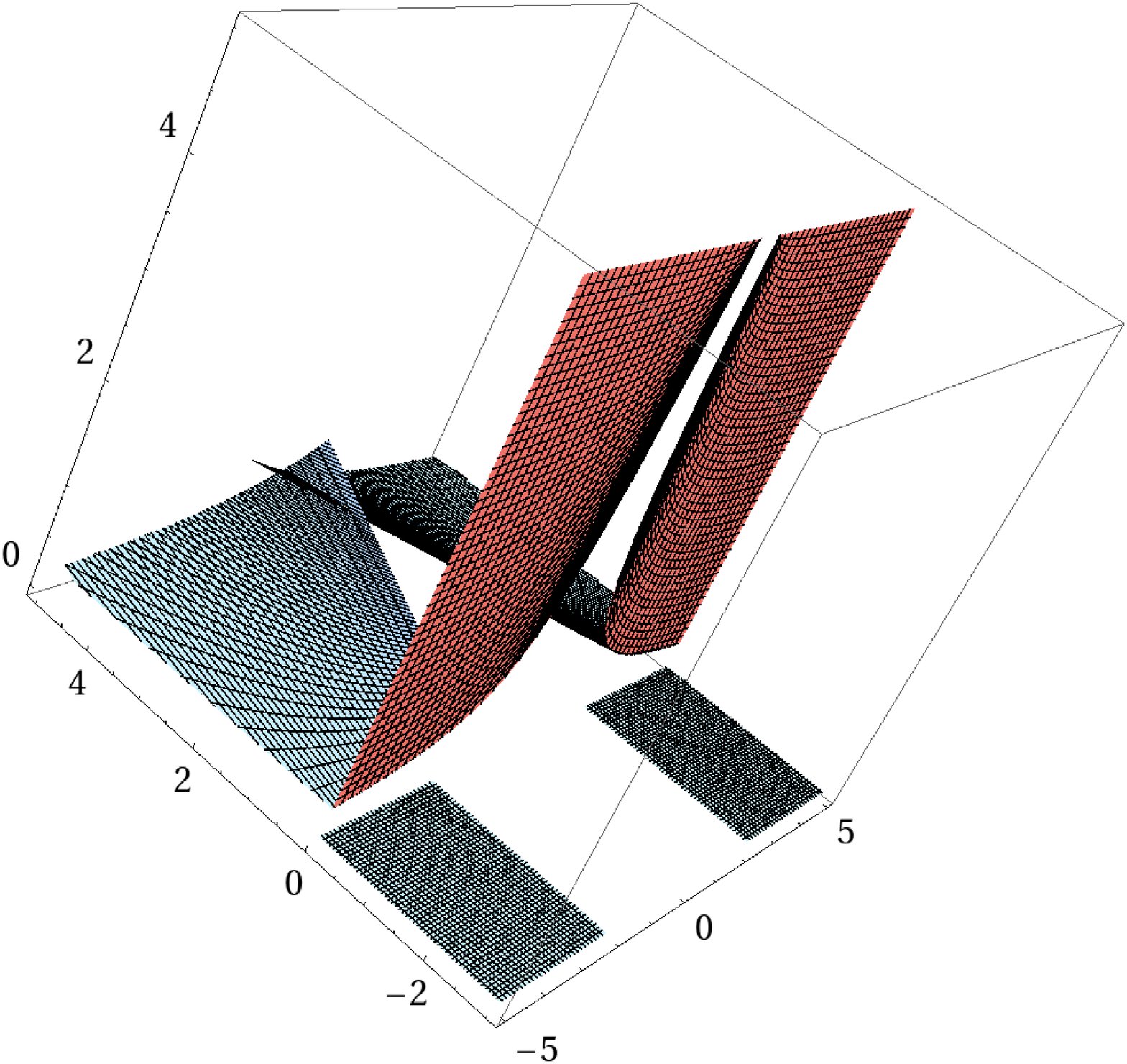}\includegraphics[width=0.33\linewidth]{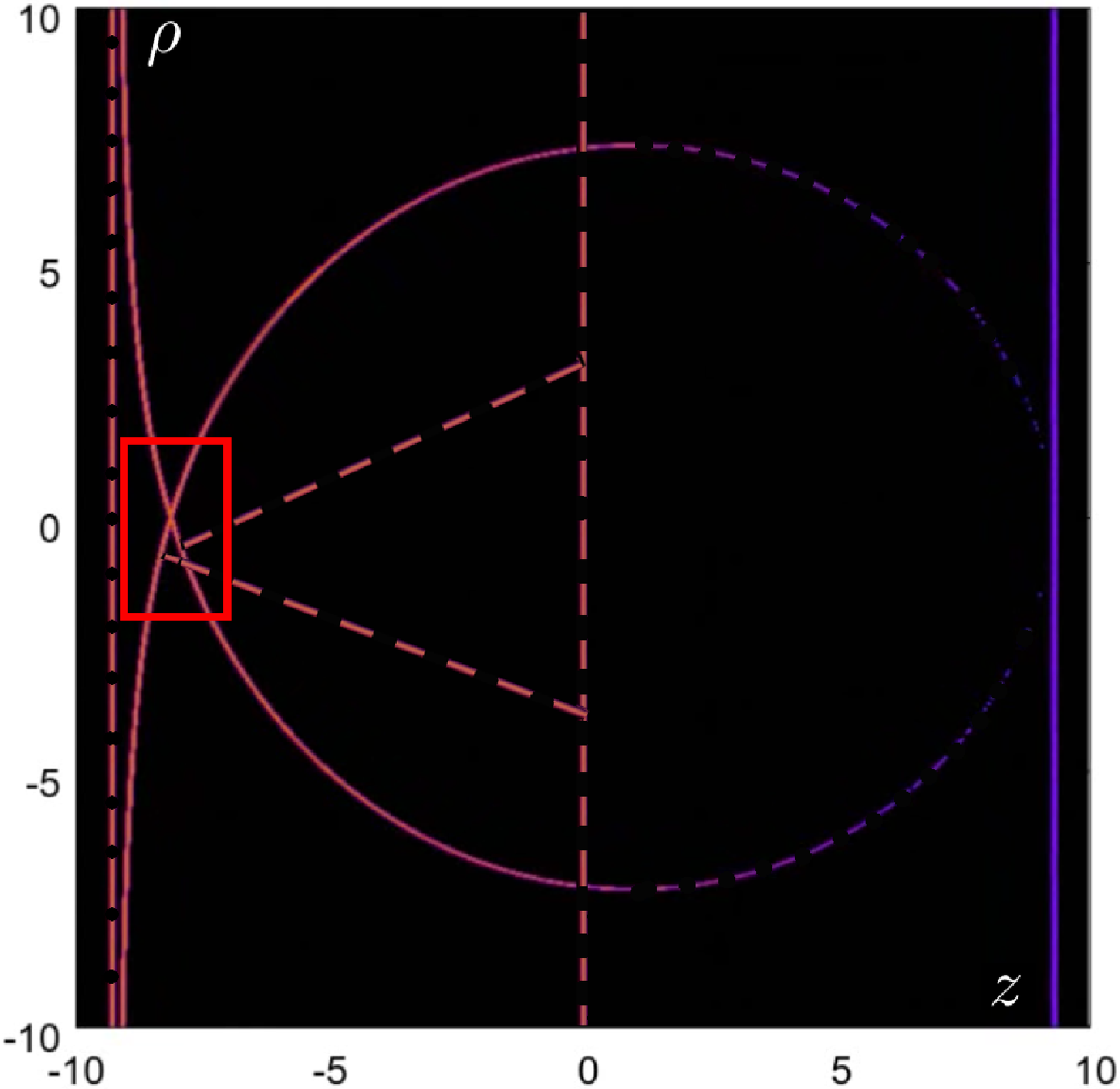}
\caption{\label{Validity} Function $g(\rho)$ which controls the validity of the perturbative approximation (left), space-time diagram representation of $v$-shock null generators for large $\rho$ (middle), and spatial slice showing the observation region behind the collision event reached by such large $\rho$ rays (right) where we expect a good perturbative approximation. Middle and right diagrams are for $D=5$.}
\end{figure}
As seen in the left panel of Fig.~\ref{Validity} the function $g(\rho)$ grows very rapidly with $\rho$ for $\rho>1$. Combining this with the fact that rays with $\rho_{incident}<1$ are trapped inside the apparent horizon, we expect the outside region of the black hole that forms to be well described by perturbation theory. In particular, if we trace the large $\rho$ rays going through the perturbative region (see middle and right panels of Fig.~\ref{Validity}), we observe that they reach an observer far away from the centre of the collision, close to the $-z$ axis of symmetry. As a consequence, the near axis metric will receive rays which are coming from large $\rho$ in the initial conditions (where the gravitational field is weak), so the closer to the axis, the better the perturbative approximation. 

\subsection{General setup and formal solution}
Now that we have identified a region of space-time that we expect to be described by a perturbation of flat Minkowski space, we are ready to use some general tools of perturbation theory applied to our initial value problem. Our criterion is that $h(v,\rho)\ll 1$, which can be represented as $h(v,\rho)\sim \epsilon$, for some small $\epsilon$. The general idea is to assume a perturbative ansatz\footnote{The original argument given by D'Eath and Payne\cite{D'Eath:1992hb,D'Eath:1992hd,D'Eath:1992qu} was based on boosting to a frame where the $u$-shock appears strong compared to the $v$-shock. The solution ansatz is nevertheless the same.} for $u>0$ such that
\begin{equation} 
g_{\mu\nu}=\eta_{ \mu \nu}+\sum_{i=1}^\infty h_{\mu\nu}^{(n)}\ ,
\label{eq:pertexpansion} 
\end{equation}
where $h_{\mu\nu}^{(n)}\sim O(\epsilon^n)$.\footnote{Note that if we had not rescaled the coordinates for each factor of $h(v,\rho)$ there would be a factor of $\kappa$ so we could also identify powers $\kappa^n$ to each order.} Now we insert this ansatz into the Einstein equations and equate order by order. At each order we have an evolution equation for the metric perturbation $h_{\mu\nu}^{(n)}$, with a source given by an effective energy-momentum tensor composed of perturbations of order $m<n$. In general the evolution equations, at each order, couple different components (see Ex.~\ref{ExPertExpand} below). However, it is well known that the de Donder gauge choice for perturbations of flat space, can be imposed order by order as to reduce the evolution equations to decoupled wave equations with a source. Thus, imposing the gauge condition
\begin{equation}
\label{gauge}
\bar{h}^{(n)\alpha\beta}_{\phantom{(n)\alpha\beta},\beta}=0  \ ,
\end{equation}
(where $\bar{h}^{(n)}_{\alpha\beta}\equiv h^{(n)}_{\alpha\beta}-\eta_{\alpha\beta}h^{(n)}/2$, $h^{(n)}\equiv \eta^{\alpha\beta}h^{(n)}_{\alpha\beta}$) one obtains the following tower of wave equations for each component
\begin{equation}\label{eq:nth_order_Feq}
\Box h^{(n)}_{\mu\nu}=T^{(n-1)}_{\mu\nu}\left[h_{\alpha\beta}^{(m<n)}\right] \ ,
\end{equation}
where the source on the right hand side is generated by the lower order perturbations, and can be computed explicitly order by order. Here $\Box=-2\partial_u\partial_v+\partial_i\partial^i$, and $h^{(n)\alpha \beta}$ are the metric perturbations in de Donder coordinates. For simplicity we have used the same notation for the metric perturbation in the new gauge to avoid extra labels. However the initial conditions~\eqref{InitCondRead} to be imposed on $u=0^+$ have to be found in this new gauge. This can be constructed order by order with the following transformation
\begin{equation}\label{Eq:coordTransfDeDonder}
  x^{ \mu}\rightarrow x^{\mu}+\sum_{i=1}^{+\infty} \xi^{(n)\mu}(x^{\alpha}) \ ,
\end{equation}
where the vector $\xi^{(n)\mu}(x^{\alpha})$ is to be determined order by order so that the de Donder gauge condition Eq.~\eqref{gauge} is obeyed\footnote{Again we use a simplified notation where the new de Donder coordinates have the same symbols.} (see Sect.~\ref{subsec:gaugefix} for details of the construction). 
\begin{Exercise}\label{ExPertExpand}
For this exercise you can use the {\em xPert} package of the {\em xAct} suite  (see first companion notebook\cite{notebook1}). 
\Question
Show that using a perturbative ansatz such as~\eqref{eq:pertexpansion} for a generic background metric $q_{\mu\nu}$ (not necessarily $\eta_{\mu\nu}$), the trace reversed linear perturbation in the (covariant) de Donder gauge obeys
\begin{equation}
\nabla^2\bar{h}^{(1)}_{\alpha\beta}-q_{\alpha\beta}R^{\mu\nu}\bar{h}^{(1)}_{\mu\nu}+\bar{h}^{(1)}_{\alpha\beta}R+2R_{\alpha\mu\beta\nu}\bar{h}^{(1)\mu\nu}-\bar{h}^{(1)\mu}_{\phantom{1}\alpha}R_{\beta\nu}-\bar{h}^{(1)\mu}_{\phantom{1}\beta}R_{\alpha\nu}=0
\end{equation}
where all derivatives, raising of indices and curvature tensors are relative to the background metric $q_{\alpha\beta}$. Specialise to the flat case and conclude that there is no source at linear order.
\Question
Check that for $i=2,3$, we get a similar equation with a source such that 
\begin{equation}
\nabla^2\bar{h}^{(n)}_{\alpha\beta}-q_{\alpha\beta}R^{\mu\nu}\bar{h}^{(n)}_{\mu\nu}+\bar{h}^{(n)}_{\alpha\beta}R+2R_{\alpha\mu\beta\nu}\bar{h}^{(n)\mu\nu}-\bar{h}^{(n)\mu}_{\phantom{1}\alpha}R_{\beta\nu}-\bar{h}^{(n)\mu}_{\phantom{1}\beta}R_{\alpha\nu}
=\bar{T}_{\alpha\beta}^{(n-1)}
\end{equation}
Specialise to the flat case for $n=2,3$ and show that the trace reversed effective energy-momentum tensor $\bar{T}_{\alpha\beta}^{(n-1)}$ is conserved.
\end{Exercise}

\subsubsection{Formal solution}

Eq.~\eqref{eq:nth_order_Feq}, can be solved using the Green's function method (see Theorem~6.3.1 of Ref.~\refcite{Friedlander:112411}), to obtain the following integral solution
\begin{equation} \label{eq:sol_orderbyorder}
h^{(n)}_{\mu\nu}=F.P.\int_{u'>0}d^{D}y'\, G(y,y')\left[2\delta(u')\partial_{v'}h^{(n)}_{\mu\nu}(y')+T^{(n-1)}_{\mu\nu}(y')\right] \ ,
\end{equation}
where $F.P.$ denotes the finite part of the integral, $y=\left\{u,v,x_i\right\}$ and the source only has support in $u>0$. The Green function for the wave operator $\Box$ is found in~\ref{AppGreen}. The interpretation of this formal solution turns out to be very simple. 
\begin{figure}
\begin{center}
\includegraphics[width=0.7\linewidth]{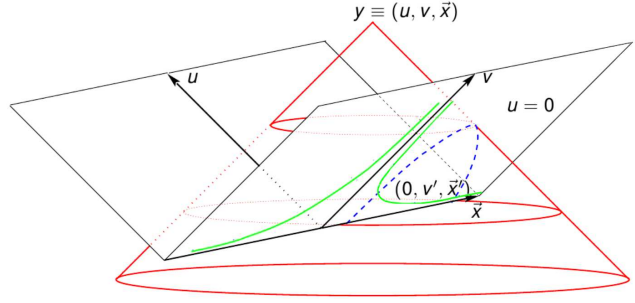}
\end{center}
\caption{\label{Fig:pastcone} (Adapted from~\cite{Herdeiro:2011ck}) Past light cone (red) of a generic space-time event $y$ after the collision, where we want to obtain the metric, and its intersection with the initial $u=0^+$ plane (blue). The collision line (for $D>4$, in green) where the $v$-shock rays jump is indicated in green. The initial data has support above the green line on such plane.}
\end{figure}
In the space-time diagram of Fig.\,\ref{Fig:pastcone} we represent the past light cone of an event $y$ after the  collision. Eq.\,\eqref{eq:sol_orderbyorder} simply states that the metric perturbation of order $n$ at the event $y$ is given by two terms which do not interfere: i) a surface term that propagates the initial data from the initial surface (compact region enclosed by the green and blue lines on $u=0^+$) to $y$, ii) and a volume term which encodes the dispersion of the wave by the background  radiation sources generated at each space-time point by lower order perturbations. Specific to the problem we are addressing, is that for $n=1$ there is no source, so the leading approximation consists of propagating the first order initial conditions in flat space. Also specific to the shock wave collision, is that introducing a second order correction, $n=2$, the initial conditions become exact. For $n>2$, there are only volume terms which encode the non-linearities of the solution.

In the remainder, we use a convenient decomposition of the metric perturbations. The following tensors on the transverse plane have already been used 
\begin{equation}\label{eq:basis_tensors}
\delta_{ij}\;,\;\Gamma_i\;,\;\Delta_{ij} \; .
\end{equation}
It is simple to check that they provide a basis of axially symmetric rank-2 tensors on the transverse plane. The metric perturbations in de Donder coordinates can then be decomposed into seven scalar functions of $(u,v,\rho)$, here denoted $A,B,C,E,F,G,H$:
\begin{eqnarray}
&h_{uu}\equiv A=A^{(1)}+A^{(2)}+\ldots \qquad &h_{ui}\equiv B \,\Gamma_i =(B^{(1)}+B^{(2)}+\ldots)\Gamma_i  \nonumber\\
&h_{uv}\equiv C=C^{(1)}+C^{(2)}+\ldots \qquad &h_{vi}\equiv F \,\Gamma_i =(F^{(1)}+F^{(2)}+\ldots)\Gamma_i \nonumber \\
&h_{vv}\equiv G=G^{(1)}+G^{(2)}+\ldots \qquad &h_{ij}\equiv E \,\Delta_{ij}+H\, \delta_{ij}\qquad\label{app:gen_perts} \\
& &\phantom{h_{ij}}= (E^{(1)}+\ldots) \Delta_{ij}+(H^{(1)}+\ldots) \delta_{ij} \ ,\qquad\nonumber 
\end{eqnarray}
where we have indicated their perturbative expansion. A similar decomposition can be applied to $T^{(n-1)}_{\mu\nu}$. With this basis, using the boundary condition on $u=0^+$ and the general solution~\eqref{eq:sol_orderbyorder}, one finds the metric perturbations by solving for these scalars after suitable contractions of~\eqref{eq:sol_orderbyorder} with the tensors~\eqref{eq:basis_tensors}.

Finally, let us mention a useful identity regarding the de Donder gauge. From the wave equations~\eqref{eq:nth_order_Feq} one can get (by reversing traces and contracting with another derivative)
\begin{equation}
\Box \bar{h}^{(n),\nu}_{\mu\nu}=\partial^\nu \bar{T}^{(n-1)}_{\mu\nu}=0\; .
\end{equation}
We have checked in Exercise~\ref{ExPertExpand} that up to third order $\bar{T}^{(n-1)}_{\mu\nu}$ is conserved, and we expect a similar result at higher orders since the background is Minkowski space and such conservation is related to the symmetries of the background. Then the formal solution for the evolution of the de Donder gauge condition is
\begin{equation} \label{eq:sol_deDonder}
\bar{h}^{(n),\nu}_{\mu\nu}=F.P.\int_{u'>0}d^{D}y'\, G(y,y')\left[2\delta(u')\partial_{v'}\bar{h}^{(n),\nu}_{\mu\nu}(y')\right] \ , 
\end{equation}
Thus the de Donder gauge condition is automatically obeyed for all $u$ if
\begin{equation}\label{deDonderV2}
\left.\partial_{v}\bar{h}^{(n),\nu}_{\mu\nu} \right|_{u=0^+}=0 \; .
\end{equation}
Thus to fix the gauge we only need to impose this condition on $u=0^+$.

\section{Detailed solutions \& Numerical evaluation}
\label{Sec:DetailedNumerics}
In this section, we apply the general results of the previous section to the initial conditions for the problem of shock wave collisions. 

\subsection{Gauge fixing}\label{subsec:gaugefix}
The basic procedure to fix the gauge order by order is as follows:
\begin{enumerate}
\item
Find the new de Donder metric perturbations, $h^{(n)}_{\mu\nu}$, in terms of the original perturbations (here denoted $h^{I(n)}_{\mu\nu}$ to avoid confusion) after the coordinate transformation~\eqref{Eq:coordTransfDeDonder} (see first notebook\cite{notebook1} for order by order expressions)
\begin{eqnarray}\label{eq:gaugeTransfOrderbyO}
h^{(1)}_{\mu\nu}&=&h^{I(1)}_{\mu\nu}+\xi^{(1)}_{\mu,\nu}+\xi^{(1)}_{\nu,\mu}\nonumber\\
h^{(2)}_{\mu\nu}&=&h^{I(2)}_{\mu\nu}+2\mathcal{L}_{\xi^{(1)}}h^{I(1)}_{\mu\nu}+\left(\mathcal{L}_{\xi^{(1)}}\mathcal{L}_{\xi^{(1)}}+\mathcal{L}_{\xi^{(2)}}\right)\eta_{\mu\nu}\\
&\ldots&\nonumber\\
h^{(n)}_{\mu\nu}&=&h^{I(n)}_{\mu\nu}+{ n\choose {n-1}}\mathcal{L}_{\xi^{(1)}}h^{I(n-1)}_{\mu\nu}+{ n\choose {n-2}}\left(\mathcal{L}_{\xi^{(1)}}\mathcal{L}_{\xi^{(1)}}+\mathcal{L}_{\xi^{(2)}}\right)h^{I(n-2)}_{\mu\nu}+\ldots \nonumber 
\end{eqnarray}
\item Insert the expression for $h^{(n)}_{\mu\nu}$, Eq.\,\eqref{eq:gaugeTransfOrderbyO} in the gauge condition Eq.~\eqref{deDonderV2}.
\item Replace the expression Eq.\,\eqref{eq:gaugeTransfOrderbyO} for $h^{(n)}_{\mu\nu}$  in the wave equation, and use it to eliminate $\partial_v\partial_uh^{I(n)}_{\mu\nu}$ derivative terms (so that only have derivatives of the initial conditions along the initial surface remain).
\item Now the gauge conditions are (schematically) in the form 
\begin{equation} \label{xiequations}
\mathcal{D}\xi^{(n)}_\mu=F_\mu(v,x_i)\leftarrow F_\mu\left[\xi^{(m<n)}_\alpha,\partial h^{I(m\leq n)}_{\alpha\beta}\right]\; ,
\end{equation}
where $\mathcal{D}$ denotes a differential operator and $F_\mu$ denotes a known function of $(v,x_i)$, which is computed at $u=0^+$ from gauge transformation vectors\footnote{These are already assumed to be computed from previous orders.} $\xi^{(m<n)}_\alpha$ and derivatives of the initial conditions along the initial surface (no $u$-derivatives).
\item Use a series expansion solution
\begin{equation}
\xi_\mu^{(n)}(u,v,x^i)=\xi_\mu^{(n,0)}(v,x^i)+u\xi_\mu^{(n,1)}(v,x^i)+\dots 
\end{equation}
to solve~\eqref{xiequations} and find the functions $\xi_\mu^{(n,0)}(v,x^i)$ and $\xi_\mu^{(n,1)}(v,x^i)$.
\item Finally use the original initial conditions $h^{I(n)}_{\mu\nu}$ and the $\xi_\mu^{(n)}$, insert in~\eqref{eq:gaugeTransfOrderbyO} and obtain explicit expression for the initial conditions in de Donder gauge.
\end{enumerate}
More detailed expressions for the gauge transformation at linear order can be found in Ref.\,\refcite{Herdeiro:2011ck}, where it is shown that the only component that transform is $h^{I(1)}_{uu}$. We will see later that the radiative components of the metric perturbations are only $h^{I(i)}_{ij}$, which at leading order are not transformed, so they are given by the second line of Eq.\,\eqref{InitCondRead}.

\subsection{Simplified integral solutions}\label{subsec:SimplifiedSols}
In the previous sections we have obtained all the pieces needed to evaluate the solutions of the shock wave problem order by order, Eq.~\eqref{eq:sol_orderbyorder}. All that remains is to: 
\begin{enumerate}
\item Contract Eq.~\eqref{eq:sol_orderbyorder} with suitable $\Gamma_i, \delta_{ij}$ or $\Delta_{ij}$ to obtain expressions for the projections~\eqref{app:gen_perts} of the metric perturbation $h^{(n)}_{\mu\nu}$ and $T^{(n-1)}_{\mu\nu}$.
\item Insert the expression for the Green function Eq.~\eqref{eqGreenFlat}. 
\item Insert the expressions for the initial data in the new gauge and, for $n\geq 2$, also the source terms.
\end{enumerate}
Though this procedure is relatively direct, the calculations to simplify the integrals are long. They involve using properties of the delta functions, integrations by parts and polynomial expansions. Here we summarise the final results of the calculations with many expressions in~\ref{AppExpressions}. 

The form of all integrals is as follows. Denote a generic (order $n$) function associated with a perturbation component of rank-$m$,\footnote{This rank is with respect to the transverse space indices $i,j$. For example $h^{(n)}_{ij}$ is rank 2 and $h^{(n)}_{iu}$ is rank 1.} by $F^{(n)}_m(u,v,\rho)$. Also denote the corresponding source function associated with the energy momentum tensor in the right hand side of the integral solution for $F^{(n)}_m$, by $S^{(n-1)}_F(u,v,\rho)$. Finally note that the initial conditions are all in the form 
\begin{equation}
F^{(n)}_m(0,v,\rho)=f(\rho)[h(v,\rho)]^n \; ,
\end{equation}
(see \ref{AppExpressions}, Eq.~\eqref{eq:frhos}, for the complete list of $f(\rho)$ functions). Then all integral solutions are written as
\begin{equation}
F^{(n)}_{m}(u,v,\rho)=F^{(n)}_{m,Surf}+F^{(n)}_{m,Vol} \; ,
\end{equation}
with
\begin{equation}
F^{(n)}_{m,Surf}=-\tfrac{n!(-1)^D\Omega_{D-4}}{(2\pi \rho)^{\frac{D-2}{2}}}\left(\tfrac{\sqrt{2}\rho}{u}\right)^n\int_{\mathcal{D}_{Surf}}d\rho'f(\rho')\rho'^{\frac{D-4}{2}+n}I_m^{D,n}(x_\star) \; ,
\end{equation}
and 
\begin{equation}\label{VolIntegrals}
F^{(n)}_{m,Vol}=-\tfrac{\Omega_{D-4}}{2(2\pi\rho)^{\frac{D-2}{2}}}\int_0^udu'\iint_{\mathcal{D}_{Vol}}dv'd\rho' \rho'^{\frac{D-4}{2}}S^{(n-1)}_{F}(u',v',\rho')I^{D,0}_m(x) \; .
\end{equation}
The domain $\mathcal{D}_{Vol}$ corresponding to the interior of the past light cone (see Fig.\,\ref{Fig:pastcone}) is defined by 
\begin{equation}\label{ConditionX}
x\equiv \frac{\rho^2+\rho'^2-2(u-u')(v-v')}{2\rho\rho'}\leq 1 \; ,
\end{equation}
and the domain $\mathcal{D}_{Surf}$, by $x_\star=x|_{\rm @ \, collision\, line}$, i.e. replacing $u'=0$ and $v'=\Phi(\rho')/\sqrt{2}$ in Eq.\,\eqref{ConditionX}. The functions $I_m^{D,n}(x)$ are defined in \ref{AppExpressions}.

\subsection{Numerical strategies for surface integrals}\label{subsec:NumericalStrateg}
In the remainder of these notes, we focus mostly on surface integrals. In fact, the surface integrals are central in the determination of the solution for the following reasons. On the one hand, the first  order perturbations are surface integrals since there is no source generated at the linear level. On the other hand,  at second order, the source function is computed from the first order metric perturbations, so it is again a contribution built from surface integrals.

Before proceeding, it is useful to re-write the surface term in the following way. Define the quantities
\begin{equation}
y\equiv\tfrac{\rho'}{ \rho} \;,\qquad p\equiv(\sqrt{2}v-\Phi(\rho))\rho^{D-4} \;,\qquad q\equiv\tfrac{u}{\rho^{D-2}}
\end{equation}
then
\begin{equation}\label{eq:Surf2Dform}
F^{(n)}_{m,Surf}=
-\frac{n!(-1)^D\Omega_{D-4}}{\rho^{(D-3)(2n+N_u-N_v)}(2\pi)^{\frac{D-2}{2}}}\left(\frac{\sqrt{2}}{q}\right)^n\int_{\mathcal{D}_{Surf}}dy \,f(y)y^{\frac{D-4}{2}+n}I_m^{D,n}(x_\star)
\end{equation}
with
\begin{equation}
x_\star=\frac{1+y^2-\sqrt{2}{q}(p-\psi(y))}{2y} \; , \; \psi(y)\equiv \left\{\begin{array}{ll} -2\log y &, D=4  \\ \frac{2}{D-4}\left[\frac{1}{y^{D-4}}-1\right] &, D>4 \end{array}\right. \; .
\end{equation}
Here $N_u,N_v$ can be thought of as the $u$-rank and the $v$-rank (i.e. the number of $u$-indices or $v$-indices) of the metric perturbation associated with $F^{(n)}_{m}$. With this simple observation, we have completely scaled out the $\rho$ dependence, and now the integral only depends on two coordinates, $(p,q)$, instead of $(u,v,\rho)$. This scaling is related to a hidden symmetry which turns out to be very useful order by order (see Sect.\,\ref{subsec:highO2D}).

The evaluation of~\eqref{eq:Surf2Dform} requires the following sensitive steps:
\begin{itemize}
\item {\em Domain determination:} The integration domain $\mathcal{D}_{Surf}$ is defined by $x_\star\leq 1$, which requires the classification of the roots of a complicated function of $y$.
\item {\em Stable implementation of $I_m^{D,n}$ functions:} There are many such functions and depending on three integer parameters. Also there are numerically difficult regions of the argument $x$ which must be implemented carefully.
\item {\em Integrable singularities of $I_m^{D,n}$:} These can be identified from the definitions in~\ref{AppExpressions}. A careful removal procedure is necessary.
\end{itemize}
We address each of these individually in the next sections.

\subsubsection{Domain determination}\label{sec:Domains}
For a given pair  $(p,q)$ (or equivalently $u,v,\rho$), corresponding to an observation point where we want to know the metric perturbations,  the condition which defines the domain of integration is simply $x_\star(y)\leq 1$. Thus $x_\star=1$ defines a boundary point in the domain of integration (see Fig.\ref{PlotDomain}). From the definitions~\eqref{app:IevenDef} and~\eqref{eq:IfuncDodd1} in appendix, clearly the points $x_\star=\pm 1$ are also singular points of the integrand in many cases, so in practice one also has to split the integrand around the $x_\star=-1$ point as well. Therefore we must study two types of domains: A) for $x_\star<-1$  and B) for $-1\leq x_{\star}\leq1$. The conditions for type-B are written in the following form ($\alpha\equiv D-4$)
\begin{equation}
\left\{ \begin{array}{c}
C_{-}(y)\equiv y^{\alpha+2}-2y^{\alpha+1}-(\sqrt{2}qp-1)y^{\alpha}+\sqrt{2}qy^{\alpha}\psi\left(y\right)\leq0\vspace{4mm}\\
C_{+}(y)\equiv y^{\alpha+2}+2y^{\alpha+1}-(\sqrt{2}qp-1)y^{\alpha}+\sqrt{2}qy^{\alpha}\psi\left(y\right)\geq 0
\end{array}\right..
\end{equation}
 To solve these conditions and obtain the domain of integration we need to classify the curves $C_{-}(y)$ and $C_{+}(y)$, which are collectively denoted
\begin{equation} 
C_{s}\equiv y^{\alpha+2}+2sy^{\alpha+1}-(\sqrt{2}qp-1)y^{\alpha}+\sqrt{2}qy^{\alpha}\psi\left(y\right)\; . 
\end{equation}
The type-A domain obeys simply the condition $C_{+}<0$.

The classification of these functions is done by noting that both are positive at $y=0$ and when $y\rightarrow +\infty$, and that they are smooth functions everywhere else. The number of roots is found by looking at the number of minima. Taking derivatives
\begin{eqnarray}
\dfrac{dC_{s}}{dy}=0&\Rightarrow& y=\dfrac{-s(\alpha+1)\pm\sqrt{1+\sqrt{2}q(\alpha+2)\left(\alpha p+2\right)}}{\alpha+2}\\
&\Rightarrow& y_{min}=\dfrac{-s(\alpha+1)+\sqrt{1+\sqrt{2}q(\alpha+2)\left(\alpha p+2\right)}}{\alpha+2}
\end{eqnarray}
 where the last step follows from the positivity of $q$ and $y$. Thus there is exactly one minimum which may or may not be negative. 

For the type-B domain, if there is a negative minimum for $C_{+}$ we just find its roots to the left and right of the minimum numerically, and exclude the region between them from the domain of integration. Otherwise if the minimum of $C_{+}$ is positive, then no restriction comes from the corresponding condition. For the curve $C_{-}$, we find the roots similarly and do the opposite, i.e. include only the region between the two zeros. The interception with the restriction from the $C_{+}$ condition gives then the type-B domain. 

For the type-A domain, the procedure is the same for the curve $C_+$ except that now we accept the region between its roots, which is the domain we seek. This picture can be checked in Fig.\,\ref{PlotDomain}.
\begin{figure}
\begin{center}
\includegraphics[width=0.9\linewidth]{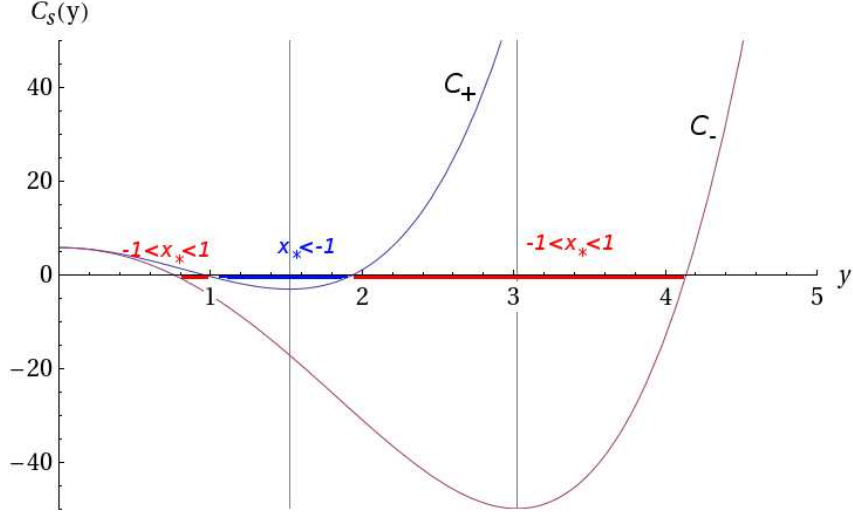}
\end{center}
\caption{\label{PlotDomain} Here we plot an example of the two curves $C_s(y)$ for $D=6$ and two positive values of $p\simeq 0.73$ and $q\simeq 4.16$. The vertical lines indicate the location of the minimum of the curves $C_+$ and $C_-$ (from left to right respectively).}
\end{figure}

Knowing that the minimum of a curve $C_s(y)$ is negative, tells us that the roots are either in $y\in \left[0,y_{\rm min}\right]$ or in $y\in \left[y_{\rm min},+\infty\right[$. For the first interval we can directly use a root bracketing algorithm which always converges to the correct solution. For the second interval, one might use a derivative algorithm since the interval is unbounded. However, derivative algorithms can be unstable and jump to wrong solutions (with negative $y$), so it is always better to find an upper bound on the location of the second root and use again a bracketing algorithm (see Exercise~\ref{ExDomain} below).
\begin{Exercise}
\label{ExDomain}
You can find solutions to these examples in the second companion notebook\cite{notebook2}.
\Question
Check that the classification of the curves $C_s(y)$ is correct by plotting several cases with different $p,q,D$ values.
\Question
Write a function which takes as arguments $p,q,D$ and returns the following cases:
\begin{itemize}
\item An empty list if there is no domain of integration available, $\{\}$.
\item An ordered list  of the roots for $C_-$ and $C_+$ (Example: $\{root_1,root_1,root_3\}$)
\item Do you think that in practice you should worry about degenerate cases such as the example in the previous line?
\end{itemize}
\end{Exercise}

\subsubsection{Integration method and $I_m^{D,n}$ functions}

The integrand of Eq.~\eqref{eq:Surf2Dform} requires a stable evaluation of the $I_m^{D,n}$ functions. As seen in~\ref{AppIeven} and~\ref{AppIodd}, many of such functions are basically polynomials. However, some involve trigonometric functions, or differences of powers of square roots with cancellations for asymptotic regions such as $|x| \gg 1$. This scaling can create numerical problems, especially when we attempt to evaluate the integrals with high precision, where finely spaced function calls are needed. The general issue of how to write expressions for numerical evaluation can be many times overlooked and create mysterious problems. A general word of advice is to be very careful on the way expressions are written and to examine (numerically) all functions in extreme regions of their arguments to investigate whether an asymptotic series representation is necessary for numerical stability. Some examples specific to the expressions for $I_m^{D,n}$ are detailed in Sect.~2.1 and~2.2 of the second companion notebook\cite{notebook2}.

Another sensitive issue is that of integrable singularities. For $n\geq0$ all integrands are either regular or have an integrable singularity, which can be removed through the following procedure. Start by re-writing 
\begin{equation}
F^{(n)}_{m,Surf}=\dfrac{n!(-1)^{D+1}\Omega_{D-4}}{(2\pi)^{\frac{D-2}{2}}}\left(\dfrac{\sqrt{2}}{q}\right)^{n}\sum_{i}\int_{\mathcal{D}_{i}}\, f(y)y^{\frac{D-4}{2}+n}\dfrac{\left(\sqrt{1-x_{\star}^{2}}I_{m}^{D,n}(x_{\star})\right)}{\sqrt{1-x_{\star}^{2}}}
\end{equation}
 where the sum is over the number of domains determined in section~\ref{sec:Domains}. Each type of domain (A and B), can be organised in sub-types, labelled by the value of $x_\star$ at the boundary point. Let us denote the ends of the domain (left/right) by $y_{\pm}^{i}$, and the corresponding $x_\star$ by $x_{\pm}^{i}$. Then for type-B, we have three sub-types: $(x_{-}^B,x_{+}^B)=(1,1),(1,-1)$ or $(-1,1)$. For type-A, there is only one possibility  $(x_{-}^A,x_{+}^A)=(-1,-1)$. The general strategy is to split each domain $\mathcal{D}_i$ in two around the middle
point $(y_{-}^{i}+y_{+}^{i})/2$, and use a change of variables adapted
to each integrable singularity at each end point $x_{\pm}^{i}$. Since
$(x_{k}^{i})^{2}=1$ ($k=\pm$) we can always write the singularity as 
\begin{equation}
\dfrac{1}{\sqrt{1-x_{\star}^{2}}}=\dfrac{1}{\sqrt{1-x_{k}^{i}x_{\star}}}\dfrac{1}{\sqrt{1+x_{k}^{i}x_{\star}}}
\end{equation}
so that the singular term (the first), is manifest at $x_{\star}=x_{k}^{i}$. Finally we perform a change of variable in the corresponding half interval as to eliminate the integrable singularity 
\begin{equation}
y=y_{k}^{i}-k\, u^{2} \; ,
\end{equation}
 and then 
\begin{multline}\label{eq:RegIntegrand}
F^{(n)}_{m,Surf}=\tfrac{n!(-1)^{D+1}\Omega_{D-4}}{(2\pi)^{\frac{D-2}{2}}}\left(\tfrac{\sqrt{2}}{q}\right)^{n}\sum_{i}2\sum_{k=\pm}\int_{0}^{\sqrt{\frac{y_{+}^i-y_{-}^i}{2}}}du\, f(y)y^{D-4+n}\times\\ \times\dfrac{\left(\sqrt{(-1)^{\delta}(1-x_{\star}^{2})}I_{m}^{D,n}(x_{\star})\right)}{\sqrt{y^{D-4}(-1)^{\delta}\frac{1-x_{k}^{i}x_{\star}}{u^{2}}}\sqrt{1+x_{k}^{i}x_{\star}}} \; ,
\end{multline}
($\delta=0$ when $x_{\star}>-1$ otherwise $\delta=1$). We have re-written the integrand as to display in the numerator (in parenthesis) a term which is regular everywhere. In the denominator the term under the first square root is also regular because the $1/u^2$ factor must cancel the zero in $1-x_{k}^{i}x_{\star}$ at the boundary point. Simplified expressions for this term which are more suitable for numerical evaluation are presented in~\ref{AppSings}. 

Later, we will need to compute derivatives of the metric functions. Note that one can write schematically
\begin{equation}
F^{(n)}_{m,Surf}=X^{(n)}\,\mathcal{K}^{(n)}\left[I_{m}^{D,n}(x_{\star})\right] \; ,
\end{equation}
 where the $\mathcal{K}^{(n)}$ operator does not contain any dependence
on $(p,q)$, and $X^{(n)}$ denotes the pre-factor outside the integral. Then, noting that $\left(I_{m}^{D,n}\right)'=-I_{m}^{D,n-1}$
\begin{eqnarray}
\partial_{\alpha}F^{(n)}_{m,Surf}& = & \partial_{\alpha}\log X^{(n)}\, X^{(n)}\mathcal{K}^{(n)}\left[I_{m}^{D,n}\right]+X^{(n-1)}\mathcal{K}^{(n-1)}\left[\tfrac{-n\sqrt{2}y}{q}\partial_{\alpha}x_{\star}I_{m}^{D,n-1}\right]\nonumber\\
 & \equiv & \sum_{j=n-1}^{n}X^{(j)}\mathcal{K}^{(j)}\left[A_{\alpha}^{(j)}I_{m}^{D,j}(x_{\star})\right] \; .
\end{eqnarray}
 Applying this expression we obtain (all others vanish) 
\begin{equation}
A_{q}^{(n)} =  -\dfrac{n}{q}\:,\qquad A_{q}^{(n-1)}=\dfrac{n(p-\psi(y))}{q}\:,\qquad A_{p}^{(n-1)}  =  n \; .
\end{equation}
\begin{Exercise}
Use the results in this section to address the following problems (see second notebook\cite{notebook2} Sect.\,2.3 and Sect.\,3)
\Question
Eq.~\eqref{eq:Fakesing4D} for $D=4$ contains $\log$ terms, so strictly speaking the singularity has not been removed explicitly. 
\begin{itemize}
\item Plot such terms and investigate what happens near the singular point. 
\item Investigate solving the problem by using a series expansion and devise a criterion to use it in the (numerically) problematic region.
\item Write a numerically stable function to return the removed singularity under the square root term, for all $D$.
\end{itemize}
\Question
Collect all the results of this section and write a wave form function which computes $E_{,p}$ for even $D$. Test with the $D=4$ case and recover plots similar to Fig.\,\ref{2D4plots}.
\end{Exercise}

\subsection{First order results and radiation extraction}\label{subsect:FirstOradExtract}

In this section we are finally ready to evaluate the first order integrals to obtain wave forms at leading order.
\begin{figure}
\includegraphics[width=\linewidth]{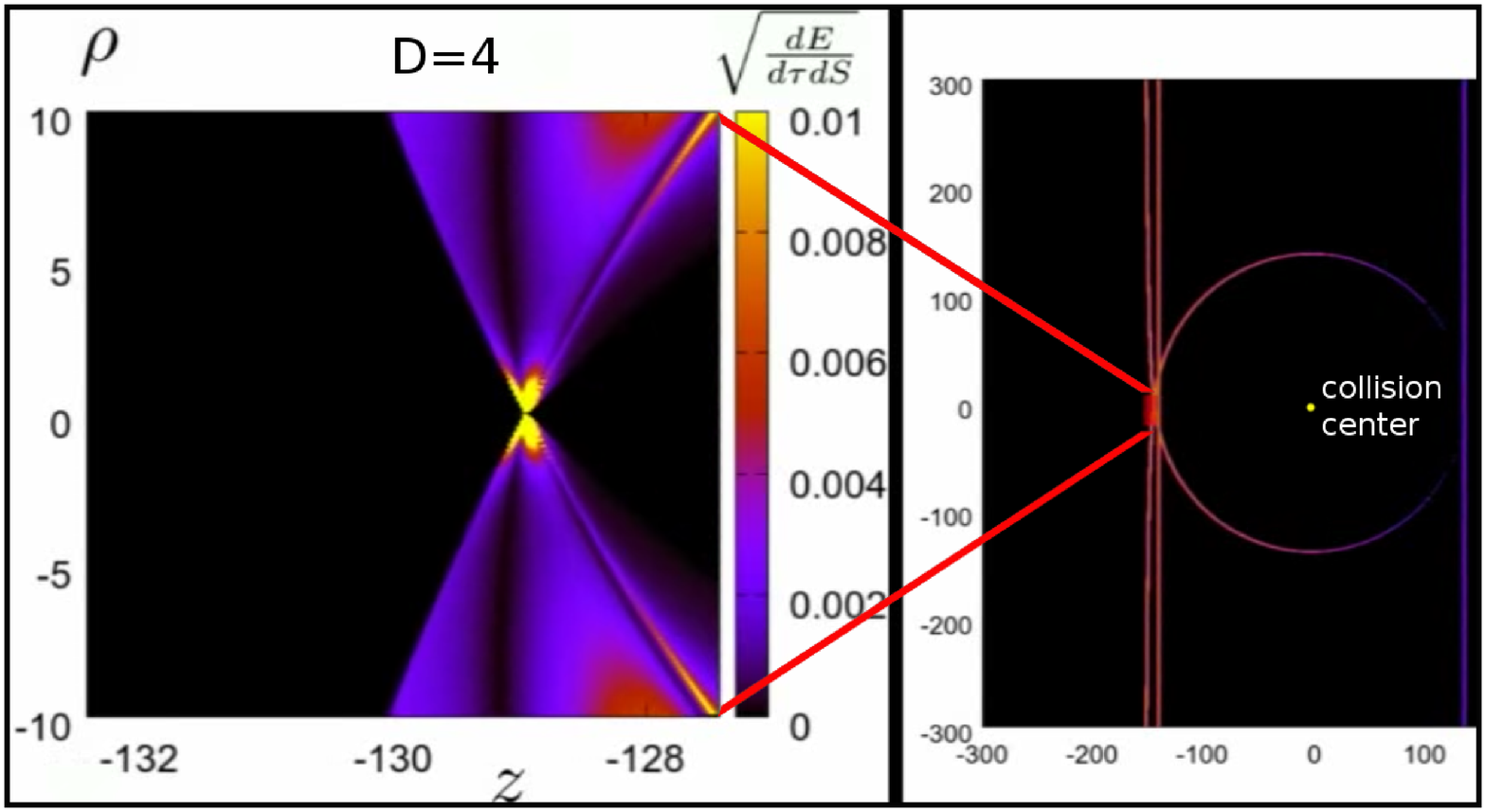}
\includegraphics[width=\linewidth]{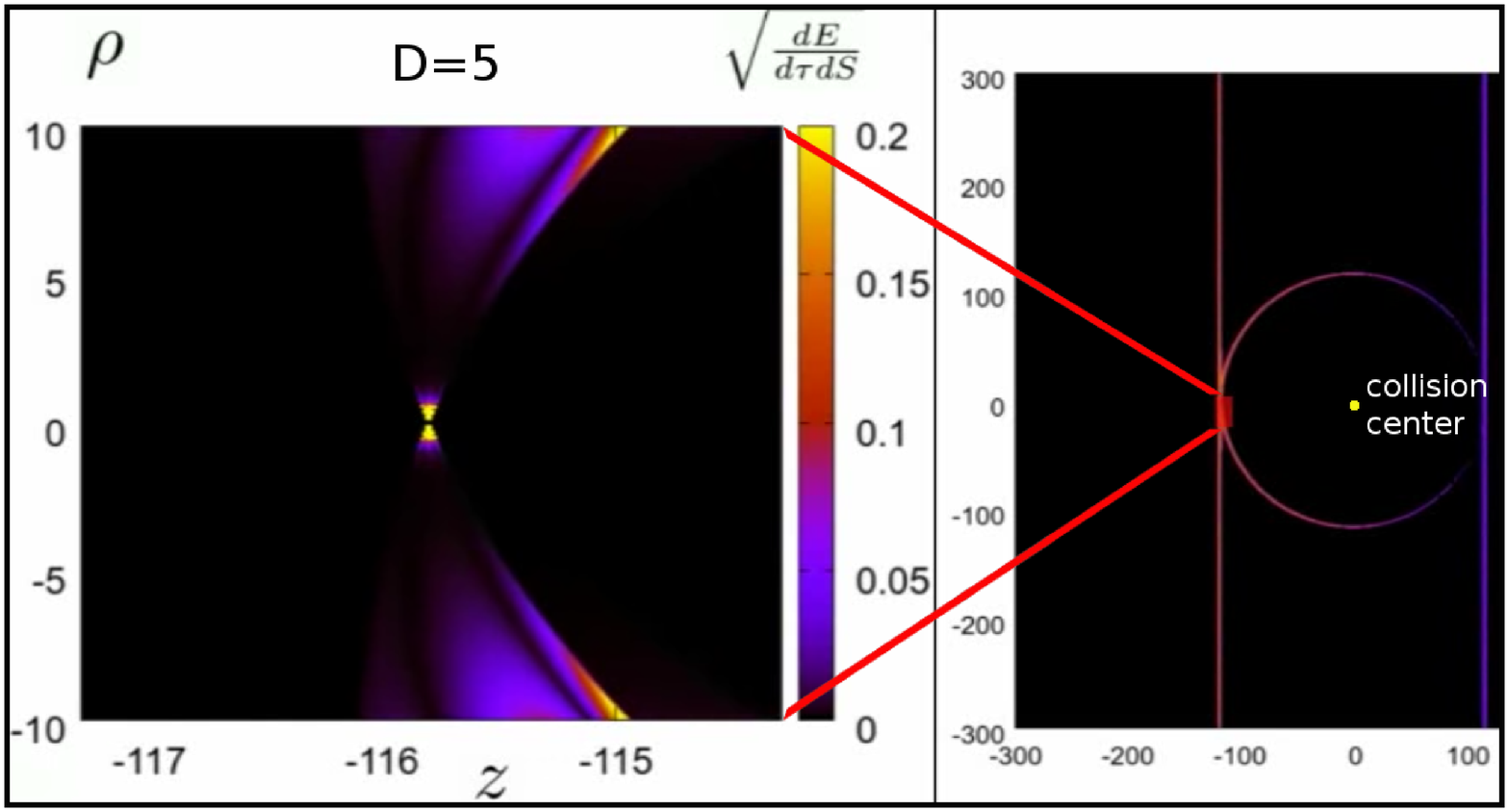}
\caption{\label{Fig:Spanshots4D5D} Snapshots of the first order approximation for the wave form signal seen for observers near the collision axis, for $D=4$ (top) and $D=5$ (bottom). Each snapshot contains, on the right, a representation of the null rays of the $u$ and $v$ shocks, where the collision centre is indicated (yellow dot) as well as the location (red rectangle) of the region which is zoomed and displayed in the left panel.}
\end{figure}
In Fig.\,\ref{Fig:Spanshots4D5D} we represent two snapshots of the magnitude of the wave form (basically the square root of the power flux -- see discussion in Sect.~\ref{subsub:RadExtract}) for $D=4$ and $D=5$ and an observation region far away from the collision centre, near the $z$-axis. The radiation signal follows closely the optical rays (see right panels), within the outermost wave front of first null rays and the circular region defined by the second optical rays (recall Fig.~\ref{FigRaysMovie1_5D}), where the signal decays towards the centre. This suppression is much sharper for odd $D$ as seen in the bottom panel. 
\begin{figure}
\includegraphics[width=0.5\linewidth]{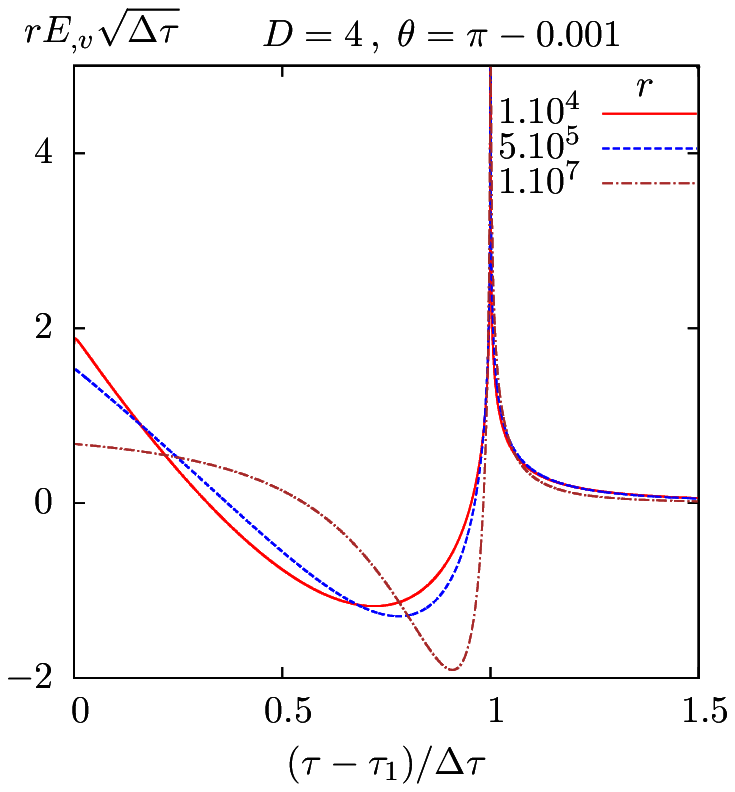}\includegraphics[width=0.5\linewidth]{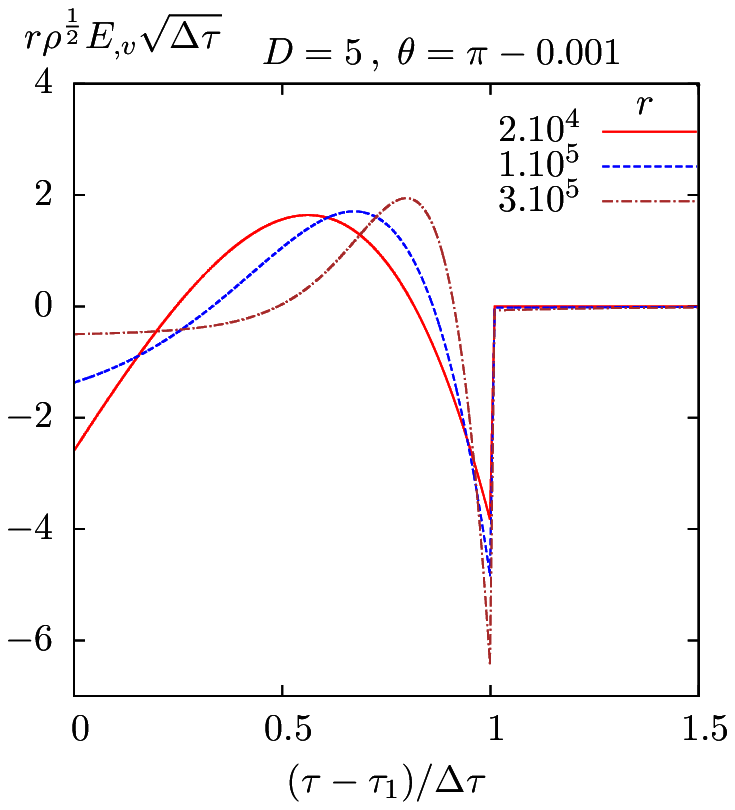}
\caption{\label{Fig:WaveForms} (Adapted from\cite{Herdeiro:2011ck,Coelho:2012sya}) Wave forms for $D=4$ and $5$ for fixed $\theta$, near the axis, and various $r$. The initial step (where the signal starts) is exactly at $\tau=\tau_1$ (time of arrival of first optical ray), whereas the sharp peak is at $\tau=\tau_2$ (second optical ray). $\Delta \tau=\tau_2-\tau_1$.}
\end{figure}
These properties are seen more clearly in Fig.~\ref{Fig:WaveForms} where wave forms seen by an observer at fixed $r,\theta$ (small angles and large radii) are plotted as a function of retarded time, $\tau\equiv t-r$.

\subsubsection{Approximations}
Now that we have found an approximation for the gravitational wave, it remains to integrate the power flux of the signal to obtain the energy that is radiated away. We have seen that the perturbative method gives a reliable wave form in a region close to the collision axis, i.e. for $\theta$ close to $\pi$ ($x\equiv \cos\theta \sim -1$). Thus, to obtain an estimate, an extrapolation  off the axis is required.

 In Refs.\,\refcite{D'Eath:1992hb,D'Eath:1992hd,D'Eath:1992qu}, D'Eath and Payne have assumed an approximation to extrapolate off the axis as follows. First, observe that our system is axially symmetric, and invariant under reflections $z\leftrightarrow -z$ so we can define an angular emission function
 \begin{equation}
C(x)\equiv \dfrac{2}{E_{CM}}\dfrac{dE}{dx}=\sum_{n=0}^{+\infty} C_n(x^2-1)^n
\end{equation}
which must be even due to the reflection symmetry which. This justifies the series expansion on the right hand side. This function is basically the news function referred to in Ref.\,\refcite{D'Eath:1992hb,D'Eath:1992hd,D'Eath:1992qu}. Then the inelasticity is 
\begin{equation}
\epsilon_{\rm radiated}=  \int_{-1}^{1} \tfrac{dx}{2}\sum_{n=0}^{+\infty}C_n(x^2-1)^{n}=\sum_{n=0}^{+\infty}\dfrac{C_n(-2)^nn!}{(2n+1)!!}\label{epsilonexpandee} \ .
\end{equation} 
If the news function $C(x)$ is analytic, the expansion close to the axis is indeed sufficient, provided that $\lim_{n\rightarrow +\infty}|C_{n+1}/C_n|\leq 1$. Furthermore, since there is an extra $n$ dependent suppression factor in~\eqref{epsilonexpandee}, we expect higher orders to become increasingly less important. The approximation used by D'Eath and Payne\cite{D'Eath:1992qu} in  $D=4$, corresponds to a truncation at second order in perturbation theory, which provides the first angular correction (truncation at $n=1$ in~\eqref{epsilonexpandee}). Their result is $\epsilon_{\rm radiated}=0.163$. This is in agreement with the latest numerical relativity simulations of ultra-relativistic particle or black hole collisions at large boost\cite{East:2012mb,Sperhake:2008ga}, so it seems to indicate that the angular corrections ($C_n$ for $n>1$) are small in $D=4$.  If instead we truncate at leading order ($n=0$) we obtain an isotropic approximation.

Regarding the quality of the truncated approximation, we can make some estimates in $D=4$. Assuming the $C_n$'s to be of the same order, we would estimate the $n=1$ correction to be $\sim 2/3$ of the leading order, by looking at Eq.\,\eqref{epsilonexpandee}. In $D=4$, we known that at leading order $C_0\simeq 0.250$ and the corrected result\cite{D'Eath:1992qu} up to order $n=1$ is $C_0-2C_1/3=0.163$. So we get that $C_1/C_0\simeq 0.52$, which implies that the second order correction is, in this case, actually less than the crude estimate of $2/3$ of the leading order.

\subsubsection{Radiation extraction methods}\label{subsub:RadExtract}
In this section we present two methods to define the news function $C(x)$ associated to the angular energy emission into gravitational waves.

The first method works at linear order and it consists of the direct computation of the radiated power using the Landau-Lifshitz pseudo-tensor\cite{landau}, which was generalised to higher dimensions in Ref.\,\refcite{Yoshino:2009xp}. In fact, such tensor is basically the source that is generated by the linear perturbations appearing on the right hand side of Eq.~\eqref{eq:nth_order_Feq} at second order. Its expression in terms of the perturbations in de Donder gauge $h_{\mu\nu}^{(1)}(u,v,x_i)$ which are traceless reads (we omit here the superscript $(1)$ for notational simplicity)
\begin{equation}
\begin{array}{rcl}
16\pi G_Dt^{\mu\nu}_{LL} & = & \displaystyle{h^{\mu\nu}_{\ \ , \alpha}h^{\alpha\beta}_{\ \ , \beta}-h^{\mu\alpha}_{\ \ , \alpha}h^{\nu\beta}_{\ \ , \beta}+\frac{1}{2}\eta^{\mu\nu}\left(h^{\alpha\beta}_{\ \ , \sigma}h^\sigma_{\ \alpha,\beta}-\frac{1}{2}h^{\beta\sigma,\alpha}h_{\beta\sigma,\alpha}\right)} \\
& & \displaystyle{-h^{\mu\beta}_{\ \ , \sigma}h_{\beta}^{\  \sigma, \nu}-h^{\nu\beta}_{\ \ , \sigma}h_{\beta}^{\  \sigma, \mu}+h^{\mu\alpha,\beta}h^\nu_{\ \alpha,\beta}+\frac{1}{2}h^{\beta\sigma,\mu}h_{\beta\sigma}^{\ \ ,\nu}} \ . \label{LL}
\end{array}
\end{equation}
Despite not being unique or gauge-invariant it is well known that the integral
\begin{equation}
E_{\rm radiated}=\int t^{0i}_{LL}n_idSdt \ , 
\end{equation}
computed on a `distant' surface with area element $dS$ outward unit normal $n^i$, is a gauge-invariant well defined energy\cite{wald}.  Using this, in Ref.\,\refcite{Herdeiro:2011ck} it was shown that under the isotropic approximation,
\begin{equation}\label{approxLL}
\epsilon_{\rm radiated}\simeq  \frac{1}{8}\dfrac{D-2}{D-3}\lim_{\hat{\theta}\rightarrow 0,r\rightarrow \infty}\left(\int (r\rho^{\frac{D-4}{2}}E_{,v}^{(1)})^2 dt \right)\; .
\end{equation}

\begin{figure}
$\phantom{.\qquad.\qquad.\qquad.}$\includegraphics[width=0.8\linewidth]{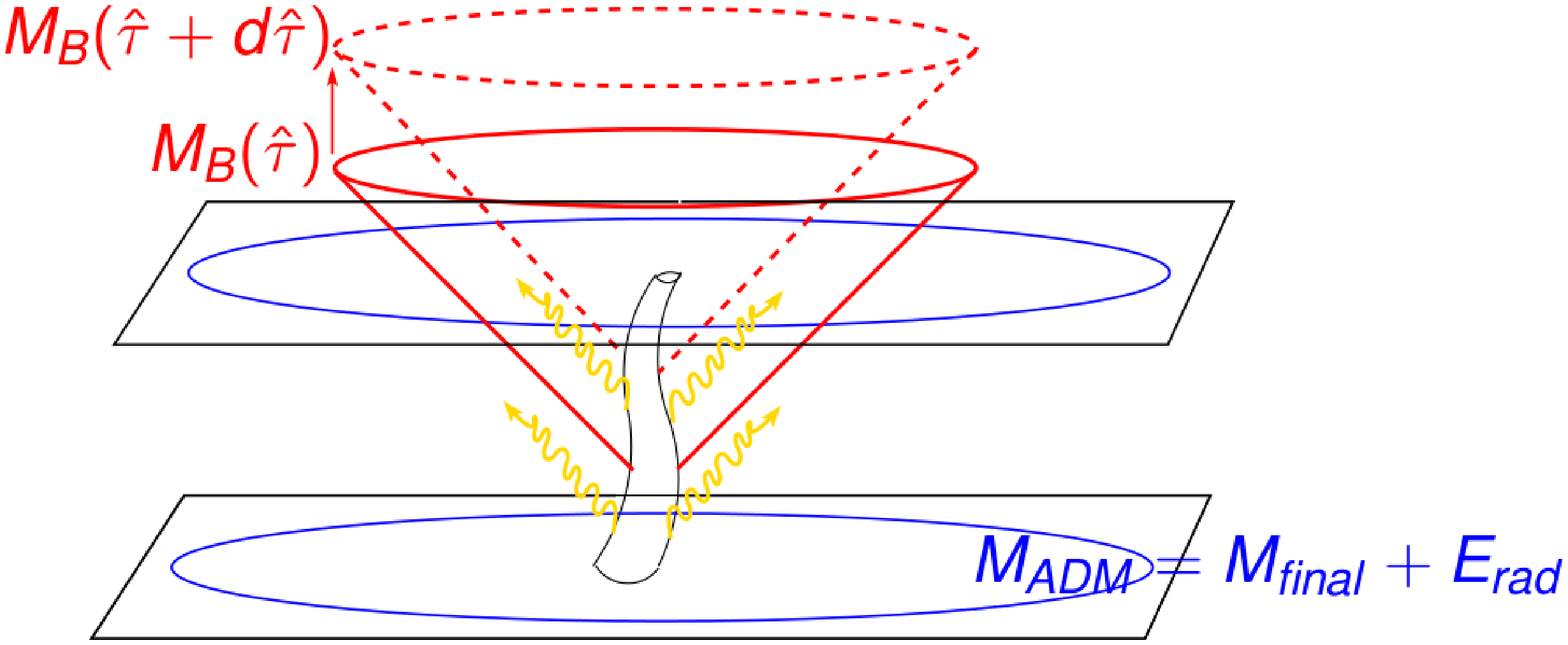}
\caption{\label{BondiADM} Schematic representation of slice where the Bondi Mass  is computed (red null slices) in contrast with the ADM mass (spatial blue slice).}
\end{figure}

A second, more general method, consists of changing coordinates to a Bondi gauge. In a Bondi gauge, the radiated power is obtained from the time variation of the Bondi mass. This can be understood pictorially in Fig.~\ref{BondiADM} where we contrast with the ADM mass.
Consider a space-time with collapsing matter and an outgoing radiation signal. The ADM mass is found on a spatial slice, at a very large radius where space-time is asymptotically flat. The ADM mass thus picks up on each slice the energy of the gravitational radiation, as well as the mass of the collapsing body. In contrast, the Bondi mass is computed at asymptotically flat null infinity, which can be seen as the limit of bending the slice used for the ADM mass until it becomes null.  This excludes the gravitational radiation, which is emitted along null surfaces. Then the variation of the Bondi mass measures the emission of radiation by the system (see Fig.\,\ref{BondiADM} red cones). In Ref.\,\refcite{Coelho:2012sy} it has been shown on general grounds (assuming only the de Donder gauge condition, asymptotic flatness, axisymmetry and transforming to Bondi gauge) that the inelasticity at all orders is
 \begin{equation}
\epsilon_{\rm radiated}= \int_{-1}^{1} \tfrac{d\cos\theta}{2}\lim_{r\rightarrow +\infty}\int d\tau W(\tau,r,\theta)^2\equiv  \int_{-1}^{1} \tfrac{dx}{2}C(x)\label{epsilon} \ ,
\end{equation} 
with
\begin{equation}
W(\tau,r,\theta)\equiv\sqrt{\frac{(D-2)(D-3)}{8}}\, \, r\rho^\frac{D-4}{2}\left(E_{,v} +H_{,v}+E_{,u}+H_{,u}\right) \label{app:waveformDef} \ .
\end{equation}  
\begin{figure}
\includegraphics[width=0.453\linewidth]{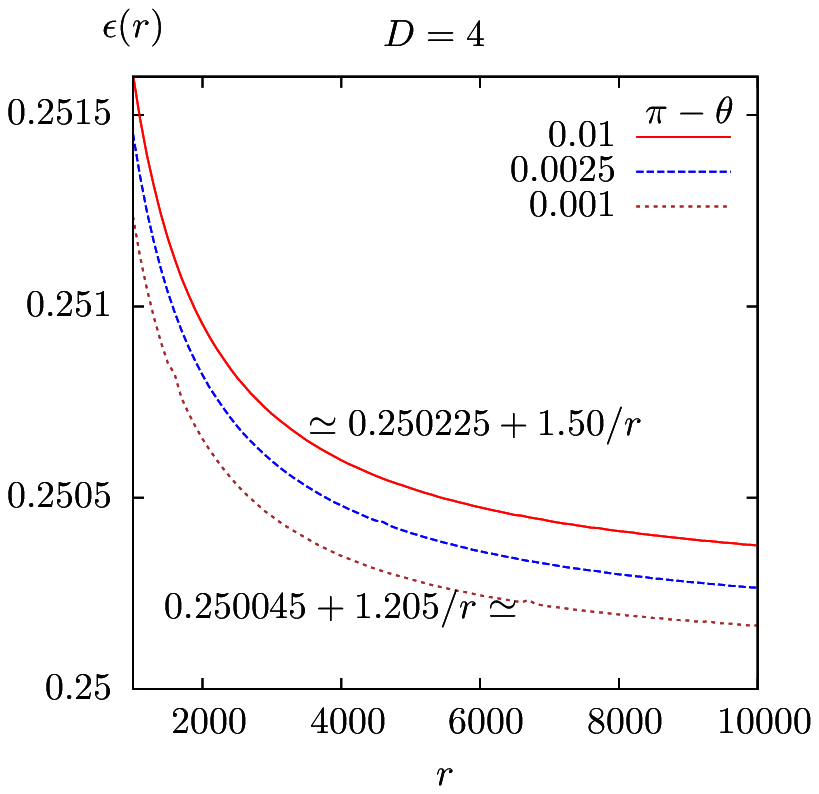}\includegraphics[width=0.547\linewidth]{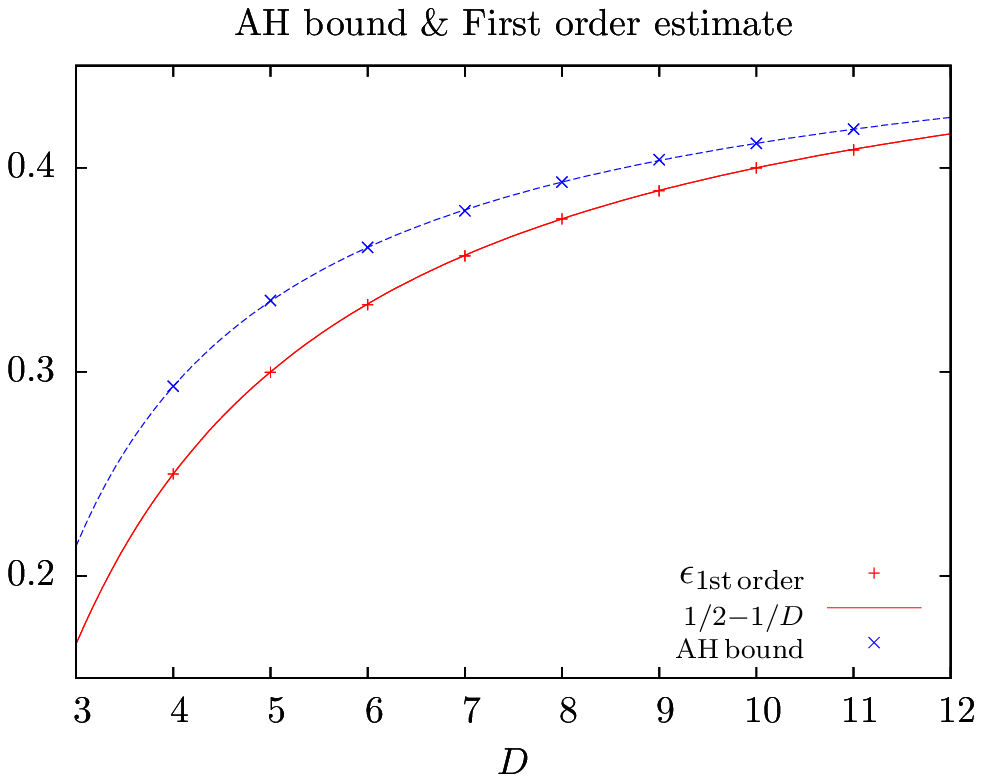}
\caption{\label{ExtractionFitFormula} (Adapted from\cite{Herdeiro:2011ck,Coelho:2012sya}) Numerical procedure to take the limit $r\rightarrow +\infty,\theta\rightarrow 0$ (left) for $D=4$, and the final first order estimate as a function of $D$ (right) compared with the apparent horizon bound. }
\end{figure}
Both methods have been used to extract the inelasticity\cite{Herdeiro:2011ck,Coelho:2012sya,Coelho:2012sy}, by numerically taking the limit $r\rightarrow +\infty$. In Fig.~\ref{ExtractionFitFormula} left, we use the $D=4$ case to illustrate the procedure. In essence, one computes the quantity under the limit Eq.~\eqref{approxLL} or~\eqref{app:waveformDef}, for very small $\theta-\pi$ and increasingly large $r$. Then a fit to a curve of the form $\epsilon_{\rm radiated}+b/r+\ldots$ is performed, and the tail is used to estimate numerical errors. 

The right panel of Fig.~\ref{ExtractionFitFormula},  shows the inelasticity computed with this approximation for various $D$, (high precision points in red). This is always within the apparent horizon bound (in blue). But most importantly there is a remarkable agreement with the following, strikingly simple, fit formula\cite{Coelho:2012sya}
\begin{equation}\label{miracle}
\epsilon_{\rm 1st \, order}=\dfrac{1}{2}-\dfrac{1}{D} \; .
\end{equation}

\subsection{Higher orders and two dimensional reduction} \label{subsec:highO2D}
In this section we briefly comment on the volume integrals which are in essence related to the non-linearities of the solution in the future of the collision (details to appear\cite{CHS2013}). The evaluation of the volume integrals~\eqref{VolIntegrals}  poses several challenges:
\begin{itemize}
\item The source function $S^{(n-1)}_{F}$ is obtained from a complicated expression in terms of the first order surface integrals~\footnote{The source is basically the Landau-Lifshitz pseudo-tensor, Eq.~\eqref{LL}.}. This requires  evaluating $\sim10$ surface terms.
\item The integration is now triple. 
\item The integration domain is given by a complicated implicit condition in two dimensions, Eq.~\eqref{ConditionX}.
\end{itemize}
By far the most constraining condition is that we have passed from a single integration (for the surface integrals) to a triple integration where we need various function calls, of functions which are themselves defined as integrals. However, D'Eath and Payne\cite{D'Eath:1992hd} have found a hidden symmetry order by order in perturbation theory, in $D=4$, which generalises for $D>4$. It consists of a boost followed by a conformal scaling as to keep the energy parameter $\kappa$ of the collision fixed. In summary\cite{D'Eath:1992hd,CHS2013} it implies the separation of the $\rho$ variable which was observed in the surface integrals, Eq.\,\eqref{eq:Surf2Dform}. Then, for example, the metric perturbations obey order by order 
\begin{equation}\label{2DmetricPerts}
h_{\mu\nu}^{(n)}(p,q,\phi_i,\rho)=\rho^{-(D-3)(2n+N_u-N_v)}f^{(n)}_{\mu\nu}(p,q,\phi_i) \; . 
\end{equation} 
Thus the corresponding volume integrals reduce to 
\begin{equation}\label{2DreducedVolInt}
f^{(n)}_{m,Vol}(p,q)=\tfrac{(-1)^{D}}{2}\iint dq'dp'{\sigma^{(n-1)}(p',q')}\, G^{(n,m)}(p,q,p',q')
\end{equation}
where we have used the two dimensional Green functions defined by
\begin{equation}\label{2DGreen}
G^{(n,m)}(p,q,p',q')\equiv \tfrac{(-1)^{D+1}\Omega_{D-4}}{(2\pi)^{\frac{D-2}{2}}}\int_{\mathcal{D}'_{surf}}dy\, y{}^{\frac{D-4}{2}-2n(D-3)}I_{m}^{D,0}(x) \; .
\end{equation}
The domain condition for $\mathcal{D}'_{surf}$ is now
\begin{equation}
x=\frac{1+y^2-\sqrt{2}(q-q'y^{D-2})(p-p'y^{-(D-4)}-\psi(y))}{2y} \leq 1 \; .
\end{equation}
This result tremendously simplifies the numerical problem because, effectively, we have reduced the triple integral to a double integral, at the expense of evaluating only one more surface integral Eq.\,\eqref{2DGreen} -- the 2-dimensional Green functions. These are just generalised versions of the first order surface integrals. The 2-dimensional source $\sigma^{(n-1)}(p',q')$ is defined in a similar way as the 2-dimensional metric perturbations, Eq.\,\eqref{2DmetricPerts}.

This result shows that the whole evolution problem can in fact be represented in a 2-dimensional form on the $(p,q)$ plane. 
\begin{figure}
\includegraphics[width=\linewidth]{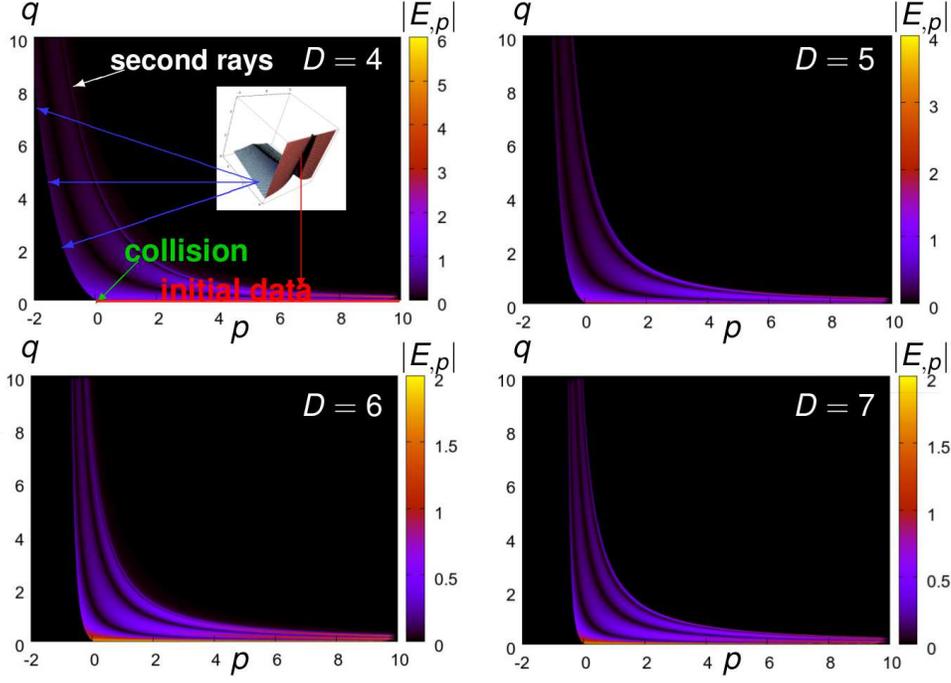}
\caption{\label{2D4plots} Magnitude of the wave form on the $(p,q)$ plane, for various $D$.}
\end{figure}
In Fig.\,\ref{2D4plots} we show plots of the magnitude of the first order wave form on such plane. In the top left panel (for $D=4$) we indicate where the initial conditions have support, the location of the first and second null optical rays and a mapping of some lines to the surfaces in the space-time diagrams of Fig.\,\ref{5DspacetimeDiag}. In this representation it becomes even clearer that the gravitational wave signal has support in the region between the two optical rays. It is also easy to observe that the number of oscillations increases with $D$. 

\begin{figure}
\includegraphics[width=\linewidth]{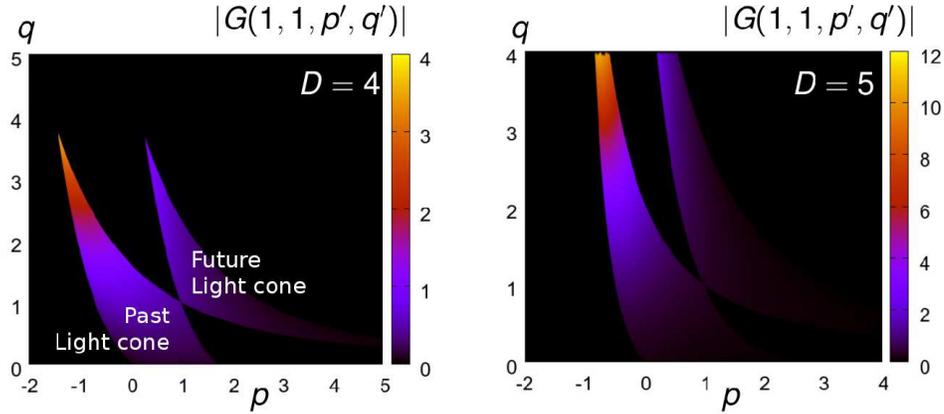}
\caption{\label{Green2Dplots} 2-dimensional Green functions on the $(p,q)$ plane for a fixed observation point.}
\end{figure}
In this representation, the volume integral, Eq.~\eqref{2DreducedVolInt}, simply encodes dispersion due to the 2-dimensional source at each point in the $(p',q')$, which is propagated with the 2-dimensional Green function, Eq.~\eqref{2DGreen}, to the observation point $(p,q)$. In Fig.~\ref{Green2Dplots}, we show the magnitude of the Green function inside the past (and future) light cone of an observation point $(p,q)=(1,1)$ for $D=4,5$ as examples. These show that the light cones on this plane have a complicated shape not at right angles. 

\section{Final remarks}
\label{Sect:FinalRemarks}

In these lecture notes, we have described in detail the problem of setting up the evolution after the collision of two AS shock waves as to model the gravitational collision of two massless particles. 

We have seen that it is possible to formulate an evolution problem with exact initial conditions by using a gauge adapted to one of the shock waves. Then, we showed how to formulate a constructive perturbative expansion to find an approximate solution for the highly curved region to the future of the collision, order by order. The discussion of the perturbative methods, namely the determination of sources, gauge fixing and some of the strategies to organise the final integral solutions, was done with relative generality, so that some results can be applied to other problems.  

In the last sections, we have discussed in detail the evaluation of surface integrals and commented on radiation extraction methods as well as on a hidden symmetry valid order by order, which seems to reduce this particular problem to a 2-dimensional problem. We have discussed the properties of the gravitational wave signal and the strikingly simple result for the inelasticity in a first order approximation, Eq.\,\eqref{miracle}. This underlying simplicity of the problem order by order in perturbation theory, opens questions about whether there is some (yet unknown) way of approaching it which allows to solve it, perhaps, non-perturbatively. The evaluation of the volume integrals to obtain a second order approximation will tell whether such a simple (or regular) structure remains at higher orders\cite{CHS2013}.

\section*{Acknowledgments}

I start by thanking the organisers of the NRHEP2 school for the support in delivering these lectures and for the stimulating environment they have provided during the school. I thank Fl\'avio Coelho and Carlos Herdeiro for their intensive collaboration in the set of projects that led to these lectures. I also thank Jai Grover for assistance and suggestions on the usage of packages in the {\em xAct} suite. M.S. is funded by FCT through the grant SFRH/BPD/ 69971/2010. This work is also supported by the grants {\em NRHEP--295189}, FP7-PEOPLE-2011-IRSES and PTDC/FIS/116625/2010.

\appendix

\section{Details of the infinite boost limit}
\label{AppBoostLimit}

 Since we expect a distribution similar to the delta function, it is natural to take the limit on an integral of the distribution. Thus we define
{\allowdisplaybreaks
\begin{eqnarray}
 F_\beta(t,z,\rho)&=&\int_{-\infty}^{z}dz'\dfrac{1}{\sqrt{1-\beta^2}\left[\left(\tfrac{z'-\beta t}{\sqrt{1-\beta^2}}\right)^2+\rho^2\right]^{\frac{D-3}{2}}} \\
&=&\dfrac{1}{\rho^{D-4}}\int_{-\infty}^{X_\beta(z,t)}dX'\dfrac{1}{\left[X'^2+1\right]^{\frac{D-3}{2}}}
\end{eqnarray} }
with
\begin{equation}
X_\beta(z,t)\equiv \frac{z-\beta t}{\rho \sqrt{1-\beta^2}} \Rightarrow \lim_{\beta\rightarrow 1}X_\beta(z,t)=sign(z-t)\times \infty
\end{equation}
With these integration limits, we obtain
\begin{equation}
\lim_{\beta\rightarrow 1}F_\beta(t,z,\rho)
=\dfrac{\theta(z-t)}{\rho^{D-4}}\int_{-\infty}^{+\infty}dX'\dfrac{1}{\left[X'^2+1\right]^{\frac{D-3}{2}}}=\dfrac{\theta(z-t)}{\rho^{D-4}}\sqrt{\pi}\dfrac{\Gamma\left(\tfrac{D-4}{2}\right)}{\Gamma\left(\tfrac{D-3}{2}\right)}
\end{equation} 
Now finally\footnote{We have used the formula $\Omega_{D-2}=2\pi^{(D-1)/2}/\Gamma(\tfrac{D-1}{2})$ and properties of the $\Gamma$ function to simplify the result.}
\begin{equation}
\lim_{\beta\rightarrow 1}\frac{4(D-2)}{(D-3)}\frac{A}{1-\beta^2}=\dfrac{16\pi G_D E}{(D-3)\Omega_{D-2}}\lim_{\beta\rightarrow 1}\dfrac{dF_\beta}{dz}=\dfrac{8\pi G_D E}{\Omega_{D-3}}\Phi(\rho)\delta(t-z)
\end{equation}

\section{Green functions}
\label{AppGreen}

The flat space Green function for the wave operator $\Box$, obeys $G(y,y')=G(y-y')$ where
\begin{equation}\label{eqGreenFlat}
G(x)=-\dfrac{1}{2\pi^{\frac{D-2}{2}}}\delta^{\left(\frac{D-4}{2}\right)}(-x^\mu x_\mu) \; .
\end{equation}
The superscript denotes the distributional derivative of order $\frac{D-4}{2}$, of the Dirac delta distribution. For odd $D$, first one defines the negative order of the delta distribution, ($p<0$) as
\begin{equation}
\delta^{(p)}(x)=\dfrac{1}{\Gamma(-p)}x^{-p-1}\theta(x) \; ,
\end{equation}
and then extend (recursively) to positive orders, by acting with derivatives on $\delta^{(p)}(x)$. For example, the fractional derivative of the delta function of order $1/2$, is 
\begin{eqnarray}\label{eq:delta12}
\delta^{(1/2)}(x)&=&\dfrac{d}{dx}\delta^{(-1/2)}(x)=\dfrac{d}{dx}\left[\dfrac{1}{\Gamma(\frac{1}{2})}|x|^{-1/2}\theta(x)\right]=-\dfrac{1}{\sqrt{\pi|x|}}\left[\dfrac{\theta(x)}{2x}-\delta(x)\right] . \qquad
\end{eqnarray}

\section{Expressions for final integrals}
\label{AppExpressions}

The initial conditions on $u=0^+$ in de Donder gauge are described by the following functions\footnote{The extra $n=2$ term for $A^{(1)}$ is due to the gauge transformation.}
{\allowdisplaybreaks
\begin{eqnarray}
A^{(1)}&:&\qquad n=1,\, f(\rho)=\dfrac{4(D-3)}{\rho^{3D-8}}\; ;\; n=2,\, f(\rho)=-\dfrac{(D-3)(D-2)}{\rho^{2D-4}} \nonumber\\
B^{(1)}&:&\qquad n=1,\, f(\rho)=\dfrac{2\sqrt{2}(D-3)}{\rho^{2D-5}}\nonumber\\
E^{(1)}&:&\qquad n=1,\, f(\rho)=-\frac{2}{\rho^{D-2}} \label{eq:frhos}\\
H^{(2)}&:&\qquad n=2,\, f(\rho)=\dfrac{(D-3)}{\rho^{2D-4}} \nonumber \\
E^{(2)}&:&\qquad n=2,\, f(\rho)=-\dfrac{(D-4)}{\rho^{2D-4}} \; \; .\nonumber
\end{eqnarray}}

The $I_m^{D,n}(x)$ functions have to be discussed separately for $D$ even or odd. In both cases the following scalars appear, related to the rank-$m$
\begin{equation}
\kappa_{0}=1,\;\kappa_{1}=-\dfrac{1}{D-3},\;\kappa_{2}=\dfrac{1}{(D-1)(D-3)} \; .
\end{equation}

\subsection{Functions $I_m^{D,n}(x)$, for $D$ even}\label{AppIeven}

\begin{equation}\label{app:IevenDef}
I_{m}^{D,n}(x)\equiv\kappa_{m}(-1)^{\frac{D-4}{2}-n}(1-x^{2})^{n-\frac{1}{2}}Q_{m}^{D,n}(x)\theta(|x|\leq1)\;,\;\; D-4\geq2n
\end{equation}
 where
\begin{equation}\label{eq:QsaDefinition}
Q_{m}^{D,n}(x)\equiv\dfrac{d_{x}^{(\frac{D-4}{2}-n+m)}\left[(1-x^{2})^{\frac{D-5}{2}+m}\right]}{(1-x^{2})^{n-\frac{1}{2}}}=\sum_{k=0}^{\frac{D-4}{2}+m-n}a_{k}x^{k}
\end{equation}
For $D-4<2n$ we have the following list of cases that are necessary
{\allowdisplaybreaks
\begin{eqnarray}
I_{0}^{4,1}(x)/\kappa_{0} & = & \pi\theta(x<-1)+\left[\dfrac{\pi}{2}-\arcsin(x)\right]\theta(|x|\leq1)\nonumber \\
I_{1}^{4,1}(x)/\kappa_{1} & = & -\sqrt{1-x^{2}}\theta(|x|\leq1)\nonumber \\
I_{2}^{4,1}(x)/\kappa_{2} & = & 3x\sqrt{1-x^{2}}\theta(|x|\leq1)\nonumber \\
I_{0}^{4,2}(x)/\kappa_{0} & = & -x\pi\theta(x<-1)+\left[x\left(\arcsin(x)-\frac{\pi}{2}\right)+\sqrt{1-x^{2}}\right]\theta(|x|\leq1)\nonumber \\
I_{1}^{4,2}(x)/\kappa_{1} & = & -\dfrac{\pi}{2}\theta(x<-1)-\left[\dfrac{\pi}{4}-\dfrac{1}{2}\left(\arcsin(x)+x\sqrt{1-x^{2}}\right)\right]\theta(|x|\leq1)\nonumber \\
I_{2}^{4,2}(x)/\kappa_{2} & = & (1-x^{2})^{\frac{3}{2}}\theta(|x|\leq1)\nonumber \\
I_{0}^{6,2}(x)/\kappa_{0} & = & \dfrac{\pi}{2}\theta(x<-1)+\left[\dfrac{\pi}{4}-\dfrac{1}{2}\left(\arcsin(x)+x\sqrt{1-x^{2}}\right)\right]\theta(|x|\leq1)\nonumber \\
I_{1}^{6,2}(x)/\kappa_{1} & = & -(1-x^{2})^{\frac{3}{2}}\theta(|x|\leq1) \nonumber\\
I_{2}^{6,2}(x)/\kappa_{2} & = & 5x(1-x^{2})^{\frac{3}{2}}\theta(|x|\leq1) \nonumber
\end{eqnarray}
}

\subsection{Functions $I_m^{D,n}(x)$, for $D$ odd}\label{AppIodd}
For $D-5\geq2n$ and $n\ge0$ 
\begin{multline}\label{eq:IfuncDodd1}
I_{m}^{D,n}(x)  =  \dfrac{\kappa_{m}}{2\sqrt{\pi}}(-1)^{\frac{D-3}{2}-n}\sum_{k=0}^{\frac{D-5}{2}+m-n}b_{k}\sum_{j=0}^{k}\binom{k}{j}\dfrac{x^{k-j}}{j-\frac{1}{2}}\left[(1-x)^{j-\frac{1}{2}}\theta(|x|\leq1)+\right.\\
  \left.+\left[(1-x)^{j-\frac{1}{2}}-(-1-x)^{j-\frac{1}{2}}\right]\theta(x<-1)\right],\qquad D-5\geq2n
\end{multline}
 with 
\begin{equation}
P_{m}^{D,n}(x)\equiv d_{x}^{(\frac{D-5}{2}-n+m)}\left[(1-x^{2})^{\frac{D-5}{2}+m}\right]=\sum_{k=0}^{\frac{D-5}{2}+n+m}b_{k}x^{k}
\end{equation}
 Otherwise (but still with $n\ge0$)
\begin{multline}
I_{m}^{D,n}(x) = \dfrac{\kappa_{m}}{\Gamma\left(n-\frac{D-4}{2}\right)}\hspace{-2mm}\sum_{k=0}^{\frac{D-5}{2}+m}\hspace{-2mm}c_{k}\sum_{j=0}^{k}\binom{k}{j}\dfrac{x^{k-j}}{j+n-\frac{D-4}{2}}\left[(1-x)^{j+n-\frac{D-4}{2}}\theta(|x|\leq1)+\right.\\ \left.+\left[(1-x)^{j+n-\frac{D-4}{2}}-(-1-x)^{j+n-\frac{D-4}{2}}\right]\theta(x<-1)\right], \qquad D-5<2n
\end{multline}
 with 
\begin{equation}
R_{m}^{D,n}(x)\equiv d_{x}^{(m)}\left[(1-x^{2})^{\frac{D-5}{2}+m}\right]=\sum_{k=0}^{D-5+m}c_{k}x^{k}\; .
\end{equation}
For some derivatives of the metric functions, some $n=-1$ terms are actually required. For that case one obtains the same result Eq.~\eqref{eq:IfuncDodd1}, plus an extra term
\begin{eqnarray}
I_{m}^{D,-1}(x)_{extra} & = & \kappa_{m}(-1)^{\frac{D-3}{2}}\left[P_{m}^{D,0}(1)\delta^{(\frac{1}{2})}(1-x)-P_{m}^{D,0}(-1)\delta^{(\frac{1}{2})}(-1-x)\right]\qquad\\
 & = & F.P.\dfrac{\kappa_{m}(-1)^{\frac{D-3}{2}}}{2\sqrt{\pi}}P_{m}^{D,0}(1)\left[\dfrac{\theta(1-x)}{(1-x)^{\frac{3}{2}}}+(-1)^{\frac{D-5}{2}+m}\dfrac{\theta(-1-x)}{(-1-x)^{\frac{3}{2}}}\right]\; .\; \nonumber
\end{eqnarray}

\subsection{Polynomials}

One can show that for $n\geq0$ 
\begin{equation}
P_{m}^{D,n}(x)=\sum_{k=\left[\frac{p-n+1}{2}\right]}^{p\equiv\frac{D-5}{2}+m}\binom{p}{k}(-1)^{k}(2k)(2k-1)\ldots(2k-p+n+1)x^{2k-p+n}
\end{equation}
 and that 
\begin{equation}
R_{m}^{D,n}(x)=\sum_{k=\left[\frac{m+1}{2}\right]}^{p\equiv\frac{D-5}{2}+m}\binom{p}{k}(-1)^{k}(2k)(2k-1)\ldots(2k-m+1)x^{2k-m}\; .
\end{equation}
 For $n=-1$
\begin{equation}
\begin{cases}
P_{m}^{D,-1}=\sum_{k=\left[\frac{p}{2}+1\right]}^{p\equiv\frac{D-5}{2}+m}\binom{p}{k}(-1)^{k}(2k)(2k-1)\ldots(2k-p+n+1)x^{2k-p-1}\vspace{2mm} \\
P_{m}^{D,-1}=0 \;,\;\; {\rm if}\ D-7<-2m
\end{cases}
\end{equation}
Finally we derive a useful recurrence tree (for any $n$). If we define  
\begin{equation}
Q_{m}^{D,n}(x)\equiv\sum_{k=0}^{q-n}a_{k}^{(n)}x^{k}
\end{equation}
($q\equiv\tfrac{D-4}{2}+m$),  we know that 
\begin{equation}
Q_{m}^{D,q\equiv\frac{D-4}{2}+m}(x)=1 \; ,
\end{equation}
 and we can use the definition, Eq.~\eqref{eq:QsaDefinition}, of the $Q_{m}^{D,n}(x)$ to show that the following recurrence relations hold: 
\begin{eqnarray}
a_{0}^{(n-1)} & = & a_{1}^{(n)}\\
a_{k}^{(n-1)} & = & a_{k+1}^{(n)}(k+1)\theta(k<p-n\wedge p-n-1>0)-(2n-2+k)a_{k-1}^{(n)} \; . \;\;
\end{eqnarray}

\subsection{Removal of explicit singularities}\label{AppSings}
If we note that we can express 
\begin{equation}
x_{\star}=\frac{C_{-x_{k}^{i}}(y)}{2y^{\alpha+1}}+x_{k}^{i}
\end{equation}
then we can expand $y=y_{k}^{i}-k\, u^{2}$$ $ and eliminate the singularity in the square root of Eq.~\eqref{eq:RegIntegrand} explicitly to obtain: For $D=4$ 
\begin{equation}\label{eq:Fakesing4D}
\frac{1-x_{k}^{i}x_{\star}}{u^{2}}=\frac{x_{k}^{i}}{2y}\left[2ky_{k}^{i}-u^{2}-2kx_{k}^{i}+\frac{2\sqrt{2}q}{u^{2}}\left(\log(1-ku^{2}/y_{k}^{i})+ku^{2}/y_{k}^{i}\right)+\frac{2\sqrt{2}qk}{y_{k}^{i}}\right]
\end{equation}
 and for $D>4$
\begin{multline}
y^{D-4}\frac{1-x_{k}^{i}x_{\star}}{u^{2}}=\dfrac{kx_{k}^{i}}{2y}\left[\sum_{m=0}^{\alpha+1}\left(\begin{array}{c}
\alpha+2\\
m+1
\end{array}\right)\left(y_{k}^{i}\right)^{\alpha+1-m}(-ku^{2})^{m}+\right.\\
\left.-2x_{k}^{i}\sum_{m=0}^{\alpha}\left(\begin{array}{c}
\alpha+1\\
m+1
\end{array}\right)\left(y_{k}^{i}\right)^{\alpha-m}(-ku^{2})^{m}+\right.\\
\left.-\left(\sqrt{2}q\left(p+\frac{2}{D-4}\right)-1\right)\sum_{m=0}^{\alpha-1}\left(\begin{array}{c}
\alpha\\
m+1
\end{array}\right)\left(y_{k}^{i}\right)^{\alpha-1-m}(-ku^{2})^{m}\right]
\end{multline}
 where we have used the zero condition of $C_{x_{k}^{i}}(y)$ to simplify.

\bibliographystyle{ws-ijmpa}
\bibliography{sample}

\end{document}